\documentclass[10pt,twocolumn,superscriptaddress,english,10pt,prl,showpacs,floatfix,aps]{revtex4-2}

\usepackage[latin9]{inputenc}
\usepackage{amsthm}
\usepackage{amsmath}
\usepackage{amssymb}
\usepackage{esint}
\usepackage{graphicx}
\usepackage{upgreek}
\usepackage{soul}
\usepackage{xspace}
\usepackage{braket}
\usepackage[backref=true,
bookmarksnumbered=true,
bookmarks=true,
bookmarksopen=true,
colorlinks=true,
citecolor=blue,
linkcolor=blue,
anchorcolor=green,
urlcolor=blue,unicode=false]{hyperref}

\makeatletter

\newcommand{\didv}{\ensuremath{\mathrm{d}I/\mathrm{d}V}}

\begin{document}

\title{Yu-Shiba-Rusinov bands in a self-assembled kagome lattice of magnetic molecules}

\author{La\"{e}titia Farinacci}
\email{laetitia.farinacci@polytechnique.org}

\affiliation{Fachbereich Physik, Freie Universit\"at Berlin, Arnimallee 14, 14195 Berlin, Germany.}
\author{ Ga\"el Reecht}
\affiliation{Fachbereich Physik, Freie Universit\"at Berlin, Arnimallee 14, 14195 Berlin, Germany.}
\author{Felix von Oppen}
\affiliation{Dahlem Center for Complex Quantum Systems and Fachbereich Physik, Freie Universit\"at Berlin, 14195 Berlin, Germany}

\author{Katharina J. Franke} 
\affiliation{Fachbereich Physik, Freie Universit\"at Berlin, Arnimallee 14, 14195 Berlin, Germany.}

\begin{abstract}
Kagome lattices constitute versatile platforms for studying paradigmatic correlated phases. 
While molecular self-assembly of kagome structures on metallic substrates is promising, it is challenging to realize pristine kagome properties because of hybridization with the bulk degrees of freedom and modified electron-electron interactions. We suggest that a superconducting substrate offers an ideal support for a magnetic kagome lattice. Exchange coupling induces kagome-derived bands at the interface, which are protected from the bulk  by the superconducting energy gap. We realize a magnetic kagome lattice on a superconductor by depositing Fe-porphin-chloride molecules on Pb(111) and using temperature-activated de-chlorination and self-assembly. This allows us to control the formation of smaller kagome precursors and long-range ordered kagome islands. Using scanning tunneling microscopy and spectroscopy at 1.6~K, we identify Yu-Shiba-Rusinov states inside the superconducting energy gap and track their hybridization from the precursors to larger islands, where the kagome lattice induces extended YSR bands. These YSR-derived kagome bands are protected inside the superconducting energy gap, motivating further studies to resolve possible spin-liquid or Kondo-lattice-type behavior. 
\end{abstract}

\maketitle

\begin{figure*}[ht!]
    \centering
    \includegraphics[width=0.9\linewidth]{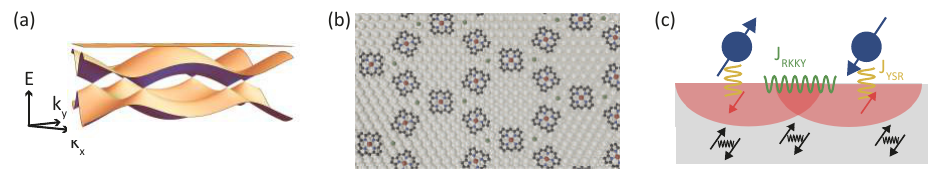}
    \caption{Kagome lattice on a superconductor. (a) Band structure of the kagome lattice with a flat band as well as the Dirac cones at the $K$ points. (b) We use molecular self-assembly of FeP-Cl molecules on Pb(111) to obtain a 2D kagome lattice. (c) At each site, exchange coupling $J_\mathrm{YSR}$ between the magnetic impurity and the substrate leads to a YSR state. Neighboring YSR states hybridize with one another. The impurity spins may also interact via RKKY interactions.
    }
    \label{fig:Fig0}
\end{figure*}

Kagome lattices are made of hexagonal tiles and corner-sharing triangles. Atomic lattices of this type feature many remarkable physical properties \cite{Yin2022}. Their band structure is characterized by strongly dispersive states with Dirac cones originating from bonding and anti-bonding states in the hexagonal lattice co-existing with a flat band resulting from self-localization of an additional electronic state (Fig.~\ref{fig:Fig0}a). Flat bands are of great general interest, as they are prone to strong many-body correlation effects \cite{Zhang2020,Si2010}, whereas the Dirac cones enable the study of relativistic effects \cite{Kang2019}.
The richness of physical phenomena increases further when spin degrees of freedom with antiferromagnetic nearest-neighbor exchange interactions enter the stage. The crystal symmetry frustrates the spin interactions, favoring spin-liquid behavior \cite{Han2012, Broholm2020}, fractional excitations \cite{Han2012}, an anomalous Hall effect \cite{Kida2011,Nakatsuji2015}, and chiral magnetic order \cite{Grohol2005}. Spin-orbit or many-body interactions may open a topological gap, expanding the class of topologically non-trivial materials \cite{Xu2015}.

So far, the study of kagome materials relied mostly on three-dimensional realizations, where stacks of kagome lattices form a bulk crystal. Inter-layer coupling alters the band structure and suppresses effects of topology and correlations depending on the coupling strength. The investigation of pristine kagome properties requires monolayers with a kagome structure. A versatile approach to growing strictly two-dimensional kagome lattices makes use of molecular self-assembly on surfaces \cite{Schlickum2008, Shi2009, Yan2021synthesis, Kumar2021}. While the first kagome lattices consisted exclusively of closed-shell molecules, inserting magnetic atoms or molecules into the lattice introduces spin in addition to the electronic degrees of freedom. 

However, despite the truly two-dimensional nature of the kagome lattice itself, its properties are stongly influenced by interactions with a bulk metallic substrate. Most notably, the substrate affects the electron-electron interactions \cite{Field2022}. We suggest that self-assembled Kagome lattices are particularly promising, when a superconducting substrate is employed. Exchange-coupled spins on superconductors induce Yu-Shiba-Rusinov (YSR) states in the superconducting gap \cite{Yu1965, Shiba1968, Rusinov1969}. States from neighboring lattice sites may hybridize \cite{Rusinov1969,Yao2014a,Ruby2018,Choi2018,Kim2018, Kezilebieke2018} and form extended bands inside the gap \cite{Pientka2013,Liebhaber2022,Schneider2022}. Hence, YSR states may be used to realize a kagome band structure, which has truly two-dimensional character and is protected by the superconducting gap. Such a system would thus provide a platform to study correlated phases of isolated kagome lattices. An even richer phase diagram emerges when the molecular spins couple (anti-ferro-)magnetically, leading to correlations between spin and electronic degrees of freedom \cite{Steiner2021} without loss of the two-dimensional character. 

A recent experiment succeeded in growing a kagome lattice of Ni atoms on Pb(111), but charge transfer quenched the magnetic moments of the Ni atoms \cite{Lin2022}. Yan \textit{et al.} \cite{Yan2021} reported self-assembly of a metal-organic kagome structure on superconducting NbSe$_2$ with kagome band formation at larger energies, but not within the superconducting energy gap. In contrast, other molecular assemblies on superconductors yielded YSR states inside the superconducting energy gap \cite{Franke2011, Farinacci2018, Homberg2020, Lu2021}, but did not feature kagome structures.

Here, we show that under appropriate conditions iron-porphin-chloride (FeP-Cl) molecules self-assemble into a kagome lattice on a superconducting Pb(111) surface (Fig.~\ref{fig:Fig0}b). We find that the molecule-induced YSR states are hybridized within smaller kagome precursors (Fig.~\ref{fig:Fig0}c) and form extended bands in larger islands. We observe edge states at kagome domain boundaries but no signatures of topologically non-trivial states. In the normal state of Pb, exchange coupling of the molecular spins to the substrate is reflected in a Kondo resonance. We suggest that kagome lattices on superconductors constitute a versatile platform for studying correlation effects owing to the protected band structure in the superconducting state while substrate-mediated magnetic interactions can be tuned, e.g., by varying the unit-cell size of the kagome lattice.

\section{Results and Discussion}

\begin{figure*}[ht!]
    \centering
    \includegraphics[width=0.9\linewidth]{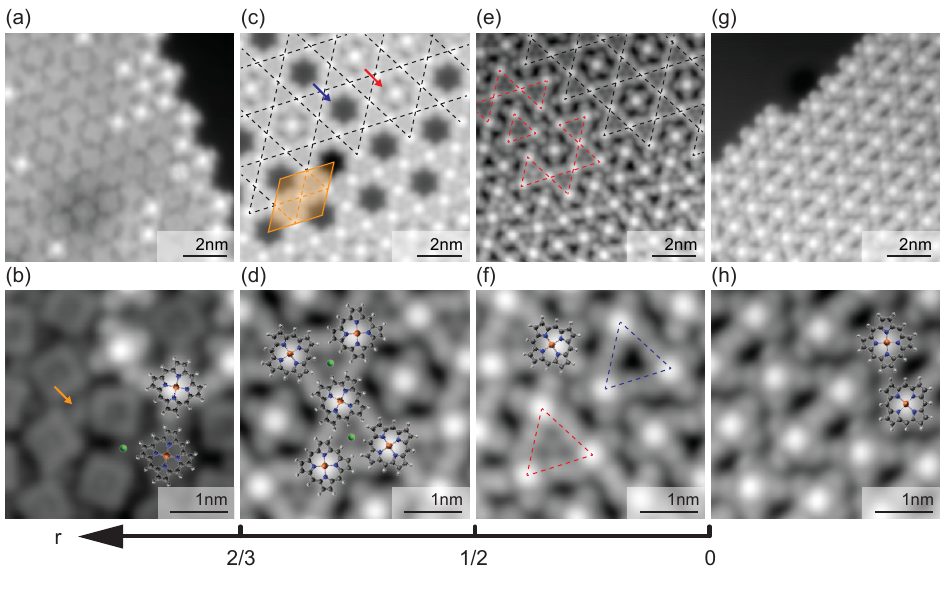}
    \caption{Kagome structure formation. (a-b) After deposition of FeP-Cl molecules on a Pb(111) sample held at $\approx$~230~K, Cl is detached from the molecules but  captured next to the FeP molecules ($V_\textrm{bias} = 20$~mV, $I=500$~pA). After annealing to $\approx$~370~K, the ratio $r$ of Cl adatoms to FeP molecules decreases, leading to the formation of (c-d) a kagome lattice for $1/2<r<2/3$  ($V_\textrm{bias} = 5$~mV, $I=50$~pA) and of (e-f) kagome precursors for $r<1/2$ ($V_\textrm{bias} = 5$~mV, $I=200$~pA). (g-h) Deposition above $\approx$~300~K leads to an hexagonal arrangement of the FeP molecules, i.e., no Cl remains on the surface ($V_\textrm{bias} = 5$~mV, $I=200$~pA).
    }
    \label{fig:Fig1}
\end{figure*}

\subsection{Self-assembled molecular structures}
We start by describing the self-assembly of Fe-porphine-chloride (FeP-Cl) molecules on Pb(111). We find that we can tune the ratio $r$ between Cl adatoms and FeP molecules such that patches of different-sized kagome precursors (Fig.~\ref{fig:Fig1}e,f) or a long-range ordered kagome lattice is formed (Fig.~\ref{fig:Fig1}c,d). As we will show later, the analysis of YSR states in the various structures allows us to track the evolution of kagome bands.

In Fig.~\ref{fig:Fig1}a, we show the molecular phase obtained after FeP-Cl deposition on a Pb(111) sample held at $\approx$~230~K. The individual molecules appear either as clover shapes with a bright protrusion at their center or as almost rectangular shapes (close-up view in Fig.~\ref{fig:Fig1}b). Some small protrusions are also observed between molecules (see arrow in Fig.~\ref{fig:Fig1}b). We surmise that these protrusions are Cl atoms that have detached from FeP-Cl molecules and that both molecular types (clover shape and rectangle) correspond to de-chlorinated FeP molecules with different electronic configurations (see section S1 in the supplementary material (SM)). The FeP molecules and Cl atoms self-assemble into a hexagonal lattice with the individual molecules being randomly oriented with respect to each other.

After annealing the low-temperature molecular phase to $\approx$~370~K, all molecules display the clover shape, and we observe the formation of a kagome lattice (Fig.~\ref{fig:Fig1}c). The lattice consists of hexagonal and triangular tiles, densely and periodically filling the surface. The triangular tiles are all occupied by a Cl atom, while the hexagonal tiles are either empty (blue arrow) or occupied by a FeP molecule (red arrow). On some parts of the same preparation we observe imperfections of the kagome lattice (Fig.~\ref{fig:Fig1}e). The periodicity of the lattice is broken into domains which still exhibit the structural motifs of triangular and hexagonal tiles (see red and black dashed lines). We will refer to these structures as kagome precursors. Not all of the triangular tiles are now occupied by Cl atoms (see red and blue triangles in Fig.~\ref{fig:Fig1}f).

The unit cell of the defect-free kagome lattice, indicated in orange in Fig.~\ref{fig:Fig1}c, consists of three FeP molecules and two Cl atoms within the triangular tiles. The ideal ratio of Cl atoms to FeP molecules is thus $r=2/3$. If $r<2/3$, the excess molecules are accommodated in the hexagonal tiles. However, once these tiles are filled at $r=1/2$, the deficit of Cl atoms can no longer be compensated and the long-range kagome lattice breaks up into smaller domains - the kagome precursors.

Deposition of the FeP-Cl molecules on a sample held above $\approx$~300~K immediately leads to all molecules being de-chlorinated as evidenced by their clover-shape (Fig.~\ref{fig:Fig1}g). The desorption barrier for Cl atoms can thus be overcome already during the adsorption process. 
The de-chlorinated molecules then assemble into large islands in which they arrange again in a hexagonal pattern with two molecular orientations, corresponding to a $45^\circ$ rotation with respect to each other. 

Concluding the structural analysis, the formation of the different molecular phases can be understood as a consequence of Cl desorption during deposition and annealing. Upon adsorption of FeP-Cl molecules on the Pb(111) substrate, the Cl ligands of the FeP molecules detach from the Fe centers. While they immediately desorb at high deposition temperature, they remain on the sample at low temperatures and are mainly captured between the molecules. The degree of Cl capture on the surface can be fine-tuned by annealing. The Cl atoms are crucial for the formation of the kagome lattice. In particular they seem to stabilize the triangular tiles. A high Cl-atom to FeP ratio $r$ ($r>2/3$) prevents the formation of the kagome lattice as there would be too many triangular tiles (eventually leading to the hexagonal lattice at $r=1$). After annealing, desorption of Cl atoms leads to a decrease of $r$ and a kagome lattice is observed for $2/3\geq r \geq 1/2$. When $r<1/2$, the excess FeP molecules can no longer be accommodated in the hexagonal tiles of the kagome lattice and long-range order of the kagome lattice is broken.

\subsection{YSR states of Kagome precursors}

\begin{figure*}[ht!]
    \centering
    \includegraphics[width=0.9\linewidth]{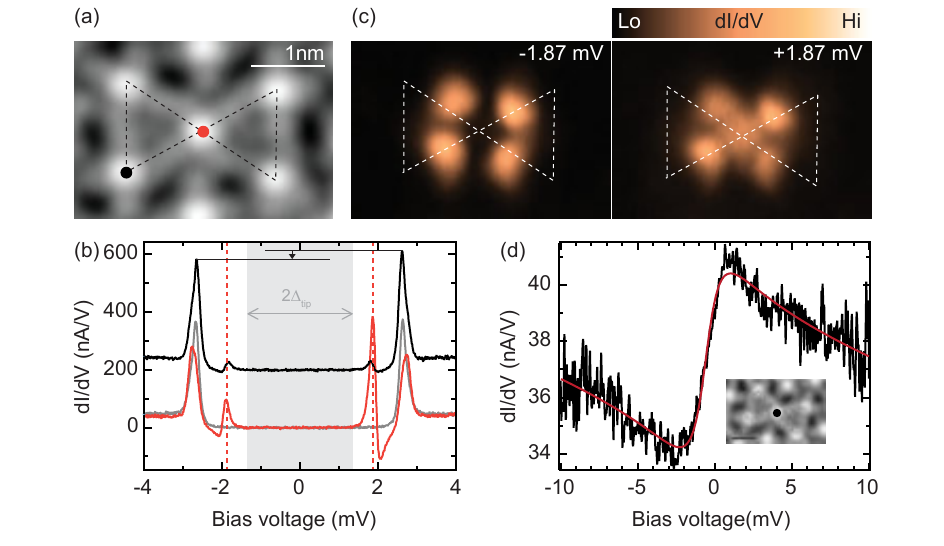}
    \caption{YSR states and Kondo resonance of the smallest kagome precursor. (a) STM topography image ($V_\textrm{bias} = 5$~mV, $I=200$~pA). (b) \didv\ spectra recorded above a molecule at the edge of the structure (black, offset for clarity) and on the central molecule (red) along with a reference spectrum taken on the bare Pb surface (grey). Both spectra are recorded on the centers of the two molecules. (c) \didv\ maps recorded at $V_\textrm{bias}=\pm 1.87$~mV while the tip follows the height profile of the STM topography image in (a) ($V_\textrm{rms}=25~\mu$eV). The dashed lines are replicas of those in (a) to help identify the molecules' positions. The \didv\ signal is present only above the central molecule. (d) Spectrum taken above the center of a similar kagome precursor and in the normal state of Pb by applying an external magnetic field $B_\mathbf{ext} = 600$~mT ($V_\textrm{rms}=50~\mu$eV). The red line is a fit to a Fano-Frota function \cite{Frank2015}, yielding a Fano width of $4.72\pm 0.23$~mV.
    }
    \label{fig:Fig2}
\end{figure*}

Next, we characterize the magnetic properties of the FeP molecules in the kagome lattice and the smaller domains of only a few kagome tiles, i.e., of the kagome precursors. In particular, we are interested in the YSR states and their evolution from smaller to larger precursors. To probe the YSR states with high energy resolution, we use superconducting Pb tips, which have been prepared by deliberately crashing the tip in the clean Pb surface. The bulk-like superconducting properties can be checked by a doubling of the superconducting energy gap in differential conductance (\didv) spectra. 

According to the analysis of the structure in Fig.~\ref{fig:Fig1}d, the two corner-sharing triangles including Cl atoms correspond to two triangular tiles of the kagome lattice. Such a kagome precursor (black dashed lines) is shown in Fig.~\ref{fig:Fig2}a. The molecule at the center of the structure (red dot) is flanked by two Cl atoms (corresponding to the occupied tiles) while the molecules at the edge of the structure only have one adatom in their vicinity. These different surroundings translate to different signatures in their differential-conductance (\didv) spectra (Fig.~\ref{fig:Fig2}b). The central molecule exhibits a pair of resonances at $\pm 1.87$~mV inside the superconducting energy gap of $\pm 2\Delta= \pm 2.7$~mV, signaling the presence of a YSR state well inside the superconducting gap. Tip-approach measurements above the Fe center further reveal that the system is in a screened spin ground state \citep{Farinacci2020} (see section S4 in SM). This is in accordance with measurements taken in an external magnetic field, which quenches superconductivity in the Pb substrate and reveals the presence of a Kondo resonance (Fig.~\ref{fig:Fig2}d). This observation highlights the quantum nature of the spin. The molecules at the edge of the structure display an asymmetry of the coherence peaks, which we attribute to a YSR state whose energy is close to the pairing energy of the substrate. The faint resonances at $\pm 1.87$~mV originate from the extended YSR state of the central molecule, as one can see in the corresponding \didv\ maps of Fig.~\ref{fig:Fig2}c. The shape and asymmetry of the YSR state in \didv\ measurements are related to interfering tunneling paths through the magnetic and frontier orbitals of the FeP molecule \cite{Farinacci2020}. Importantly, the YSR states on the center molecule are sharp and do not display indications of hybridization with other YSR states despite the close vicinity of the corner molecules. We attribute the absence of hybridization to the large energy difference of the YSR states which is presumably related to the local variations of the electronic density of states at the Fermi level \cite{Liebhaber2019, Homberg2020}. Hence, the presence of the Cl atoms is not only important for stabilizing the kagome structure but also for tuning the energy alignment of the YSR states. We note that we do not find indications of hybridization of YSR states of molecules at the edge of the structure (see section S2 in SM). However, even if there was weak hybridization, it would occur at energies well away from those that pertain to the molecules within the lattice, which we describe in the following.

\subsection{Hybridization of YSR states}

\begin{figure*}[ht!]
    \centering
    \includegraphics[width=0.9\linewidth]{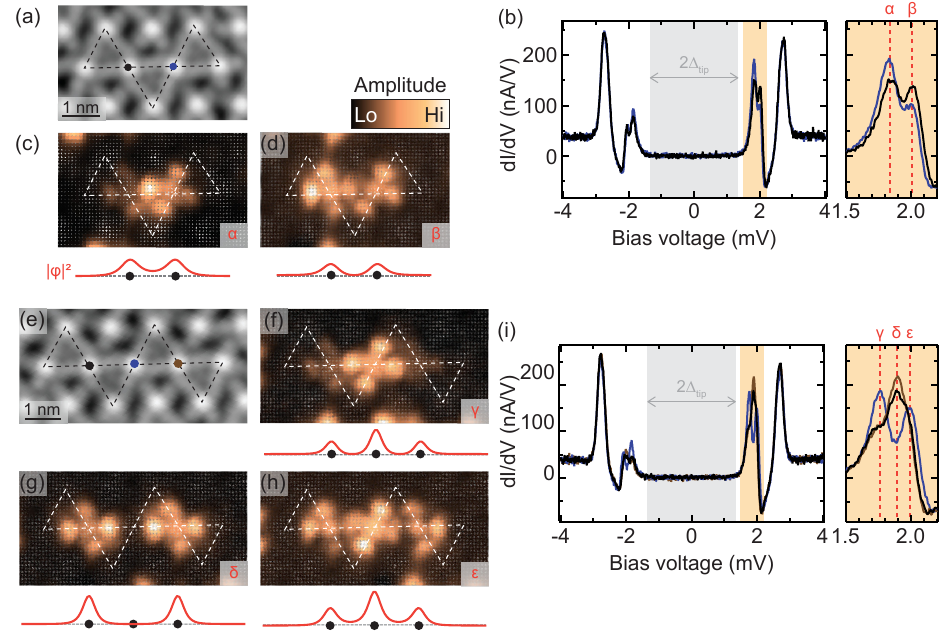}
    \caption{Hybridization of YSR states in kagome precursors. (a) Topographic image ($V_\textrm{bias} = 5$~mV, $I=200$~pA) of a kagome precursor with two molecules that have two Cl adatoms in their vicinity. (b) \didv spectra recorded above the centers of these molecules show the presence of two YSR states. The intensity of these states across the structure is obtained by recording a grid of \didv spectra ($V_\textrm{rms} = 15~\mu$eV) as shown in (c) and (d) and fitting the amplitude of each state after deconvolution (see section S3 in SM). As sketched below each data panel, the intensity distribution of the states matches that of a two-site chain with nearest neighbor coupling. (e) Topographic image ($V_\textrm{bias} = 5$~mV, $I=200$~pA) of a kagome precursor with three molecules that have two Cl adatoms in their vicinity. Three YSR states are now observed in \didv\ (i) ($V_\textrm{rms} = 15~\mu$eV) and the amplitude distributions of the states shown in (f-h) match those of a three-site chain with nearest neighbor coupling. 
    }
    \label{fig:Fig3}
\end{figure*}

Hybridization of YSR states can be observed in larger structures of triangular tiles, where each triangle shares a vertex with another one. This implies that several neighboring FeP molecules have two Cl adatoms in their vicinity and thus exhibit identical YSR states. Figure~\ref{fig:Fig3}a and e show topographic images of kagome precursors (black dashed lines) with two and three such molecules, respectively. The \didv\ spectra recorded above their Fe centers (as indicated by dots in the topographic images) show a splitting of the YSR state into two (Fig.~\ref{fig:Fig3}b) or three (Fig.~\ref{fig:Fig3}i) resonances, with intensities depending on the probed molecule. The splitting is a strong indication of YSR hybridization \cite{Flatte2000,Ruby2018,Choi2018, Liebhaber2022}. To confirm the coupling of YSR states within these structures we unravel the spatial distribution of the YSR states by recording \didv\ spectra along densely spaced grids. We remove the contribution of the tip density of state (DOS) by numerical deconvolution of the \didv\ spectra. We subsequently obtain the amplitude of each YSR state by fitting the sample DOS by a sum of Lorentzians (see section S3 in SM). The extracted amplitudes of the YSR states are then plotted in Fig.~\ref{fig:Fig3}c,d and f-h (the dashed lines serve as guide-to-the-eye by indicating the structures of the kagome precursors determined from the simultaneously recorded topographic images). 

The spatial intensity distribution along the molecular structure can be reproduced by a simple tight-binding model with nearest-neighbor coupling between sites. A sketch of the expected intensity distribution is shown below each data panel. More precisely, the kagome precursor of Figure~\ref{fig:Fig3}a shows a YSR state (labeled $\alpha$) with the largest intensity between the vertices of the triangle, i.e., between two FeP molecules (Fig.~\ref{fig:Fig3}c), whereas the YSR state labeled $\beta$ has its intensity maxima at the vertices and a nodal plane in between (Fig.~\ref{fig:Fig3}d). These YSR states thus match anti-symmetric and symmetric linear combinations of the YSR wavefunctions of the individual units (compare to Fig.~\ref{fig:Fig2}c).
Correspondingly, the kagome precursor of Fig.~\ref{fig:Fig3}e exhibits hybridized YSR states that concord with those of a three-site chain. One state is mainly localized above the central site (Fig.~\ref{fig:Fig3}f, YSR state $\gamma$), another one above the ends of the chain (Fig.~\ref{fig:Fig3}g, YSR state $\delta$), and the last one is distributed over all three sites (Fig.~\ref{fig:Fig3}h, YSR state $\epsilon$). 

These results evidence hybridization of YSR states when the smallest precursor of two corner-sharing triangles is extended to larger assemblies of triangles. To understand whether the spins associated with the YSR states can give rise to interesting magnetic properties in the extended kagome lattice, we need to search for magnetic coupling between the units. By analyzing the shift of the YSR state upon approach with the STM tip, we conclude that the ground state is a screened-spin state (see section S4 in SM). A fully screened spin would not be available for magnetic coupling. However in the gas phase, the FeP molecule carries a spin of $S=1$. If screening occurs only in one channel, the ground state remains a spin-1/2 system, which could couple via RKKY interactions. In contrast to our observations, strong RKKY coupling would lead to deviations from simple tight-binding chain behavior \cite{Steiner2021}. This suggests that in the present system, RKKY coupling is small compared to the hybridization energy.

\subsection{YSR band formation in the kagome lattice}

\begin{figure*}[ht!]
    \centering
    \includegraphics[width=0.9\linewidth]{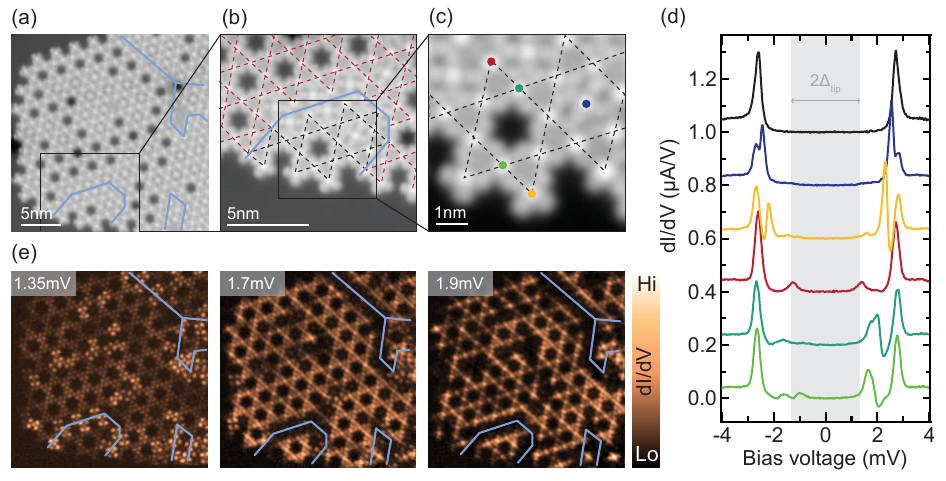}
    \caption{Kagome bands and kagome boundaries. (a) Topographic image ($V_\textrm{bias} = 5$~mV, $I=50$~pA) of a molecular island in which several kagome domains coexists, with domain boundaries indicated by blue lines. (b) Enlarged view of such a domain ($V_\textrm{bias} = -45$~mV, $I=100$~pA). (c) Molecules that have different positions relative to the kagome domains ($V_\textrm{bias} = -45$~mV, $I=100$~pA) display (d) different fingerprints in \didv\ spectroscopy ($V_\textrm{rms}=20~\mu$eV). A reference spectrum taken on bare Pb is shown in black. Spectra are offset for clarity. (e) Constant-height \didv\ maps taken at bias voltages indicated in the top left corner ($V_\textrm{rms}=25~\mu$eV).
    }
    \label{fig:Fig4}
\end{figure*}

Now that hybridization is established in kagome precursors, we examine the formation of YSR bands in a kagome lattice. The molecular island in Fig.~\ref{fig:Fig4}a exhibits several kagome domains delineated by blue lines. An enlarged view of such a domain boundary is shown in Fig.~\ref{fig:Fig4}b, where the two kagome lattices, indicated by red and black dashed lines, are mismatched by one molecular row. A close-up view of the black-dashed kagome domain is shown in Fig.~\ref{fig:Fig4}c. 
Molecules inside the kagome lattice with two neighboring adatoms (green spectra) show a pair of broad resonances between $\sim \pm 1.4$~mV and $\sim \pm 2$~mV (Fig.~\ref{fig:Fig4}d). 
In view of the hybridization of the YSR states in the kagome precursors in Fig.~\ref{fig:Fig3}, we assign this pair of broad resonances to YSR bands. In agreement with the observations above, molecules that are not surrounded by two Cl atoms exhibit sharp and energetically isolated states. For example, a molecule inside the hexagonal tile (blue spectrum) shows a YSR state close to the coherence peaks (compare to isolated molecules in Fig.~\ref{fig:Fig2}). Molecules at the edge of the lattice show either a YSR state around $\pm 1.35$~mV, corresponding to the Fermi energy of the sample (red spectrum), or close to the pairing energy (yellow spectrum), depending on the precise environment. 

To map out the YSR bands in larger kagome domains and search for possible edge states, we show \didv\ maps taken at various energies starting from the Fermi energy in Fig.~\ref{fig:Fig4}e. Interestingly, we find increased intensity along the domain boundaries of the kagome lattice inside the island (indicated by blue lines). In contrast, the domain boundaries at the edge of the island do not light up. This indicates that we do not observe a discretized edge mode along the finite boundaries of the domains, but rather an edge effect. This zero-energy feature is thus related to defects in the kagome lattice, but also potentially to local changes in the ratio between Cl adatoms and FeP molecules.  With increasing bias voltage, the enhanced intensity at the domain boundaries vanishes and the \didv\ signal is seen delocalized over the entire kagome domains. This is consistent with extended YSR bands.  

Based on the analysis of the coupling starting from the single units (Fig.~\ref{fig:Fig3}), the kagome lattice realized here involves one fermionic degree of freedom per site. We therefore expect, in the simplest case, the formation of three YSR-derived kagome bands: Two dispersive bands with a linear dispersion around the $K$ points, and a flat band (Fig.~\ref{fig:Fig0}a). In general, the \didv\ spectra include contributions from all parts of the Brillouin zone and thus do not allow for a direct identification of the individual bands or $k$-dependent band gaps. Additionally, the limited sizes of the domains as well as the irregular boundaries impede a direct analysis of the electronic dispersion by Fourier transform, as done for instance in \cite{Schneider2021}. While our experiments thus preclude the identification of a flat band, we also note that spin-orbit coupling introduces $p$-wave superconducting pairing which adds a dispersive correction to the flat bands (see section S6 in supplementary material).   

\begin{figure*}[ht!]
    \centering
    \includegraphics[width=0.9\linewidth]{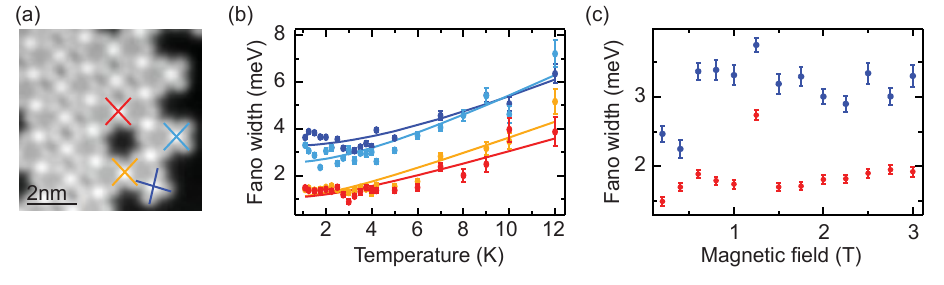}
    \caption{Evolution of Kondo resonance with increasing temperature and magnetic field. (a) Topographic image of an island featuring a kagome lattice in which we investigate the evolution of the \didv\ spectra with increasing temperature and magnetic field. Two molecules are within the island (marked by red and yellow crosses), and two molecules at the edge of the domain (indicated by light and dark blue crosses). The analysis of these spectra is the shown in (b) and (c). (b) The evolution with temperature of the width of the Kondo resonances (circles -- see section S5 in SM for more details) for these four molecules follows that of a typical Kondo system as indicated by the solid lines which are fits to $2\sqrt{(\alpha k_B T)^2+(k_B T_K)^2}$ \cite{Zhang2013, Nagaoka2002}. (c) Similarly, we do not observe any variation of the resonance widths within the range of magnetic field accessible in our experiments (see section S5 in SM for more details).
    }
    \label{fig:Fig6}
\end{figure*}

Finally, we investigate the magnetic fingerprints of the molecules when superconductivity in the underlying substrate is quenched. Molecules at the edge and within the islands (see Fig.~\ref{fig:Fig6}a) all display a Kondo resonance, albeit with different widths (see section S5 in SM for details). This is in line with the different binding energies of the YSR states indicating different coupling strengths to the substrate electrons. To reveal potential signatures of Kondo-lattice behavior we characterize the evolution of these Kondo resonances with temperature (Fig.~\ref{fig:Fig6}b) and external magnetic field (Fig.~\ref{fig:Fig6}c). The temperature-dependent measurements follow that of typical single-impurity Kondo systems, as evidenced by the good quality of the fits to $2\sqrt{(\alpha k_B T)^2+(k_B T_K)^2}$ \cite{Zhang2013, Nagaoka2002} shown as solid lines in Fig.~\ref{fig:Fig6}b. Furthermore, we do not observe a splitting or broadening of the Kondo resonance for either molecules when applying an external magnetic field up to 3T, thus giving a lower bound for their characteristic magnetic fields. In agreement with our earlier conclusion of small magnetic interactions of the YSR states, magnetic coupling towards Kondo-lattice behavior is also small and is not resolved in our experiments.

\section{Conclusions}

In conclusion, we have employed a self-assembly strategy to create a kagome lattice of spin-carrying FeP molecules and Cl atoms on a superconducting Pb(111) surface.  The role of the Cl atoms is both to stabilize the triangular tiles of the kagome lattice and to shift the YSR states of the FeP deeper into the superconducting energy gap. By varying the ratio of Cl atoms and FeP molecules, we could tune the size of the kagome domains.
This allowed us to track the hybridization of the YSR states and to infer the existence of extended YSR bands in the long-range ordered kagome lattice.

Magnetic coupling in the kagome lattice is weak, both compared to the YSR hybridization and to the individual Kondo coupling to the substrate. For future experiments, we suggest to increase the magnetic coupling by tuning the distance between the spins in the kagome lattice and using a 2D substrate. This may lead to Kondo-lattice or spin-liquid behavior in the normal state, and to exotic excitations within the superconducting gap.
 
\section*{Methods}

For all sample preparations, the Pb substrate was cleaned by sputtering with Ne$^+$ ions at 0.9~kV and annealing to 430~K under ultra-high vacuum ($\sim 10^{-10}$~mbar) conditions. The molecules were subsequently evaporated from a Knudsen cell held at 485~K. The sample was then transferred into the STM chamber where the measurements were performed at a temperature of 1.6 K. \didv\ spectra were acquired using a conventional lock-in technique and after opening the feedback at $V_\textrm{bias} = 5$~mV, $I=200$~pA. The amplitude of the voltage modulation $V_\textrm{rms}$ is indicated in the figure captions. Details concerning the deconvolution of the \didv\ spectra and subsequent fit of the sample DOS are available in section S3 of SM.

\section*{Acknowledgement}
We thank Nils Bogdanoff and Benjamin Heinrich for fruitful discussions. We thank Marie-Laure Bocquet for discussions of the molecular kagome structure. We acknowledge financial support by the Deutsche Forschungsgemeinschaft (DFG, German Research Foundation) through projects 277101999 (CRC 183, project C03) and FR2726/10-1.
%\begin{figure}[ht]
%\begin{center}
%\includegraphics{TOC.pdf}
%\end{center}
%\caption{Table of content graphic }
%\label{TOC}
%\end{figure}

%\bibliography{bibliography}

\begin{thebibliography}{40}%
\makeatletter
\providecommand \@ifxundefined [1]{%
 \@ifx{#1\undefined}
}%
\providecommand \@ifnum [1]{%
 \ifnum #1\expandafter \@firstoftwo
 \else \expandafter \@secondoftwo
 \fi
}%
\providecommand \@ifx [1]{%
 \ifx #1\expandafter \@firstoftwo
 \else \expandafter \@secondoftwo
 \fi
}%
\providecommand \natexlab [1]{#1}%
\providecommand \enquote  [1]{``#1''}%
\providecommand \bibnamefont  [1]{#1}%
\providecommand \bibfnamefont [1]{#1}%
\providecommand \citenamefont [1]{#1}%
\providecommand \href@noop [0]{\@secondoftwo}%
\providecommand \href [0]{\begingroup \@sanitize@url \@href}%
\providecommand \@href[1]{\@@startlink{#1}\@@href}%
\providecommand \@@href[1]{\endgroup#1\@@endlink}%
\providecommand \@sanitize@url [0]{\catcode `\\12\catcode `\$12\catcode
  `\&12\catcode `\#12\catcode `\^12\catcode `\_12\catcode `\%12\relax}%
\providecommand \@@startlink[1]{}%
\providecommand \@@endlink[0]{}%
\providecommand \url  [0]{\begingroup\@sanitize@url \@url }%
\providecommand \@url [1]{\endgroup\@href {#1}{\urlprefix }}%
\providecommand \urlprefix  [0]{URL }%
\providecommand \Eprint [0]{\href }%
\providecommand \doibase [0]{https://doi.org/}%
\providecommand \selectlanguage [0]{\@gobble}%
\providecommand \bibinfo  [0]{\@secondoftwo}%
\providecommand \bibfield  [0]{\@secondoftwo}%
\providecommand \translation [1]{[#1]}%
\providecommand \BibitemOpen [0]{}%
\providecommand \bibitemStop [0]{}%
\providecommand \bibitemNoStop [0]{.\EOS\space}%
\providecommand \EOS [0]{\spacefactor3000\relax}%
\providecommand \BibitemShut  [1]{\csname bibitem#1\endcsname}%
\let\auto@bib@innerbib\@empty
%</preamble>
\bibitem [{\citenamefont {Yin}\ \emph {et~al.}(2022)\citenamefont {Yin},
  \citenamefont {Lian},\ and\ \citenamefont {Hasan}}]{Yin2022}%
  \BibitemOpen
  \bibfield  {author} {\bibinfo {author} {\bibfnamefont {J.-X.}\ \bibnamefont
  {Yin}}, \bibinfo {author} {\bibfnamefont {B.}~\bibnamefont {Lian}},\ and\
  \bibinfo {author} {\bibfnamefont {M.~Z.}\ \bibnamefont {Hasan}},\ }\bibfield
  {title} {\bibinfo {title} {Topological kagome magnets and superconductors},\
  }\href {https://doi.org/10.1038/s41586-022-05516-0} {\bibfield  {journal}
  {\bibinfo  {journal} {Nature}\ }\textbf {\bibinfo {volume} {612}},\ \bibinfo
  {pages} {647} (\bibinfo {year} {2022})}\BibitemShut {NoStop}%
\bibitem [{\citenamefont {Zhang}\ \emph {et~al.}(2020)\citenamefont {Zhang},
  \citenamefont {Yin}, \citenamefont {Ikhlas}, \citenamefont {Tien},
  \citenamefont {Wang}, \citenamefont {Shumiya}, \citenamefont {Chang},
  \citenamefont {Tsirkin}, \citenamefont {Shi}, \citenamefont {Yi} \emph
  {et~al.}}]{Zhang2020}%
  \BibitemOpen
  \bibfield  {author} {\bibinfo {author} {\bibfnamefont {S.~S.}\ \bibnamefont
  {Zhang}}, \bibinfo {author} {\bibfnamefont {J.-X.}\ \bibnamefont {Yin}},
  \bibinfo {author} {\bibfnamefont {M.}~\bibnamefont {Ikhlas}}, \bibinfo
  {author} {\bibfnamefont {H.-J.}\ \bibnamefont {Tien}}, \bibinfo {author}
  {\bibfnamefont {R.}~\bibnamefont {Wang}}, \bibinfo {author} {\bibfnamefont
  {N.}~\bibnamefont {Shumiya}}, \bibinfo {author} {\bibfnamefont
  {G.}~\bibnamefont {Chang}}, \bibinfo {author} {\bibfnamefont {S.~S.}\
  \bibnamefont {Tsirkin}}, \bibinfo {author} {\bibfnamefont {Y.}~\bibnamefont
  {Shi}}, \bibinfo {author} {\bibfnamefont {C.}~\bibnamefont {Yi}}, \emph
  {et~al.},\ }\bibfield  {title} {\bibinfo {title} {Many-body resonance in a
  correlated topological kagome antiferromagnet},\ }\href
  {https://link.aps.org/doi/10.1103/PhysRevLett.125.046401} {\bibfield
  {journal} {\bibinfo  {journal} {Phys. Rev. Lett.}\ }\textbf {\bibinfo
  {volume} {125}},\ \bibinfo {pages} {046401} (\bibinfo {year}
  {2020})}\BibitemShut {NoStop}%
\bibitem [{\citenamefont {Si}\ and\ \citenamefont {Steglich}(2010)}]{Si2010}%
  \BibitemOpen
  \bibfield  {author} {\bibinfo {author} {\bibfnamefont {Q.}~\bibnamefont
  {Si}}\ and\ \bibinfo {author} {\bibfnamefont {F.}~\bibnamefont {Steglich}},\
  }\bibfield  {title} {\bibinfo {title} {Heavy fermions and quantum phase
  transitions},\ }\href {https://doi.org/10.1126/science.1191195} {\bibfield
  {journal} {\bibinfo  {journal} {Science}\ }\textbf {\bibinfo {volume}
  {329}},\ \bibinfo {pages} {1161} (\bibinfo {year} {2010})}\BibitemShut
  {NoStop}%
\bibitem [{\citenamefont {Kang}\ \emph {et~al.}(2019)\citenamefont {Kang},
  \citenamefont {Ye}, \citenamefont {Fang}, \citenamefont {You}, \citenamefont
  {Levitan}, \citenamefont {Han}, \citenamefont {Facio}, \citenamefont
  {Jozwiak}, \citenamefont {Bostwick}, \citenamefont {Rotenberg}, \citenamefont
  {Chan}, \citenamefont {McDonald}, \citenamefont {Graf}, \citenamefont
  {Kaznatcheev}, \citenamefont {Vescovo}, \citenamefont {Bell}, \citenamefont
  {Kaxiras}, \citenamefont {van~den Brink}, \citenamefont {Richter},
  \citenamefont {Ghimire}, \citenamefont {Checkelsky},\ and\ \citenamefont
  {Comin}}]{Kang2019}%
  \BibitemOpen
  \bibfield  {author} {\bibinfo {author} {\bibfnamefont {M.}~\bibnamefont
  {Kang}}, \bibinfo {author} {\bibfnamefont {L.}~\bibnamefont {Ye}}, \bibinfo
  {author} {\bibfnamefont {S.}~\bibnamefont {Fang}}, \bibinfo {author}
  {\bibfnamefont {J.-S.}\ \bibnamefont {You}}, \bibinfo {author} {\bibfnamefont
  {A.}~\bibnamefont {Levitan}}, \bibinfo {author} {\bibfnamefont
  {M.}~\bibnamefont {Han}}, \bibinfo {author} {\bibfnamefont {J.~I.}\
  \bibnamefont {Facio}}, \bibinfo {author} {\bibfnamefont {C.}~\bibnamefont
  {Jozwiak}}, \bibinfo {author} {\bibfnamefont {A.}~\bibnamefont {Bostwick}},
  \bibinfo {author} {\bibfnamefont {E.}~\bibnamefont {Rotenberg}}, \bibinfo
  {author} {\bibfnamefont {M.~K.}\ \bibnamefont {Chan}}, \bibinfo {author}
  {\bibfnamefont {R.~D.}\ \bibnamefont {McDonald}}, \bibinfo {author}
  {\bibfnamefont {D.}~\bibnamefont {Graf}}, \bibinfo {author} {\bibfnamefont
  {K.}~\bibnamefont {Kaznatcheev}}, \bibinfo {author} {\bibfnamefont
  {E.}~\bibnamefont {Vescovo}}, \bibinfo {author} {\bibfnamefont {D.~C.}\
  \bibnamefont {Bell}}, \bibinfo {author} {\bibfnamefont {E.}~\bibnamefont
  {Kaxiras}}, \bibinfo {author} {\bibfnamefont {J.}~\bibnamefont {van~den
  Brink}}, \bibinfo {author} {\bibfnamefont {M.}~\bibnamefont {Richter}},
  \bibinfo {author} {\bibfnamefont {M.~P.}\ \bibnamefont {Ghimire}}, \bibinfo
  {author} {\bibfnamefont {J.~G.}\ \bibnamefont {Checkelsky}},\ and\ \bibinfo
  {author} {\bibfnamefont {R.}~\bibnamefont {Comin}},\ }\bibfield  {title}
  {\bibinfo {title} {Dirac fermions and flat bands in the ideal kagome metal
  {FeSn}},\ }\href {https://doi.org/10.1038/s41563-019-0531-0} {\bibfield
  {journal} {\bibinfo  {journal} {Nat. Mater.}\ }\textbf {\bibinfo {volume}
  {19}},\ \bibinfo {pages} {163} (\bibinfo {year} {2019})}\BibitemShut
  {NoStop}%
\bibitem [{\citenamefont {Han}\ \emph {et~al.}(2012)\citenamefont {Han},
  \citenamefont {Helton}, \citenamefont {Chu}, \citenamefont {Nocera},
  \citenamefont {Rodriguez-Rivera}, \citenamefont {Broholm},\ and\
  \citenamefont {Lee}}]{Han2012}%
  \BibitemOpen
  \bibfield  {author} {\bibinfo {author} {\bibfnamefont {T.-H.}\ \bibnamefont
  {Han}}, \bibinfo {author} {\bibfnamefont {J.~S.}\ \bibnamefont {Helton}},
  \bibinfo {author} {\bibfnamefont {S.}~\bibnamefont {Chu}}, \bibinfo {author}
  {\bibfnamefont {D.~G.}\ \bibnamefont {Nocera}}, \bibinfo {author}
  {\bibfnamefont {J.~A.}\ \bibnamefont {Rodriguez-Rivera}}, \bibinfo {author}
  {\bibfnamefont {C.}~\bibnamefont {Broholm}},\ and\ \bibinfo {author}
  {\bibfnamefont {Y.~S.}\ \bibnamefont {Lee}},\ }\bibfield  {title} {\bibinfo
  {title} {Fractionalized excitations in the spin-liquid state of a
  kagome-lattice antiferromagnet},\ }\href
  {https://doi.org/10.1038/nature11659} {\bibfield  {journal} {\bibinfo
  {journal} {Nature}\ }\textbf {\bibinfo {volume} {492}},\ \bibinfo {pages}
  {406} (\bibinfo {year} {2012})}\BibitemShut {NoStop}%
\bibitem [{\citenamefont {Broholm}\ \emph {et~al.}(2020)\citenamefont
  {Broholm}, \citenamefont {Cava}, \citenamefont {Kivelson}, \citenamefont
  {Nocera}, \citenamefont {Norman},\ and\ \citenamefont
  {Senthil}}]{Broholm2020}%
  \BibitemOpen
  \bibfield  {author} {\bibinfo {author} {\bibfnamefont {C.}~\bibnamefont
  {Broholm}}, \bibinfo {author} {\bibfnamefont {R.~J.}\ \bibnamefont {Cava}},
  \bibinfo {author} {\bibfnamefont {S.~A.}\ \bibnamefont {Kivelson}}, \bibinfo
  {author} {\bibfnamefont {D.~G.}\ \bibnamefont {Nocera}}, \bibinfo {author}
  {\bibfnamefont {M.~R.}\ \bibnamefont {Norman}},\ and\ \bibinfo {author}
  {\bibfnamefont {T.}~\bibnamefont {Senthil}},\ }\bibfield  {title} {\bibinfo
  {title} {Quantum spin liquids},\ }\href
  {https://doi.org/10.1126/science.aay0668} {\bibfield  {journal} {\bibinfo
  {journal} {Science}\ }\textbf {\bibinfo {volume} {367}},\ \bibinfo {pages}
  {eaay0668} (\bibinfo {year} {2020})}\BibitemShut {NoStop}%
\bibitem [{\citenamefont {Kida}\ \emph {et~al.}(2011)\citenamefont {Kida},
  \citenamefont {Fenner}, \citenamefont {Dee}, \citenamefont {Terasaki},
  \citenamefont {Hagiwara},\ and\ \citenamefont {Wills}}]{Kida2011}%
  \BibitemOpen
  \bibfield  {author} {\bibinfo {author} {\bibfnamefont {T.}~\bibnamefont
  {Kida}}, \bibinfo {author} {\bibfnamefont {L.}~\bibnamefont {Fenner}},
  \bibinfo {author} {\bibfnamefont {A.}~\bibnamefont {Dee}}, \bibinfo {author}
  {\bibfnamefont {I.}~\bibnamefont {Terasaki}}, \bibinfo {author}
  {\bibfnamefont {M.}~\bibnamefont {Hagiwara}},\ and\ \bibinfo {author}
  {\bibfnamefont {A.}~\bibnamefont {Wills}},\ }\bibfield  {title} {\bibinfo
  {title} {{The giant anomalous Hall effect in the ferromagnet Fe$_3$Sn$_2$ $-$
  a frustrated kagome metal}},\ }\href
  {https://doi.org/10.1088/0953-8984/23/11/112205} {\bibfield  {journal}
  {\bibinfo  {journal} {J. Phys.: Condens. Matter}\ }\textbf {\bibinfo {volume}
  {23}},\ \bibinfo {pages} {112205} (\bibinfo {year} {2011})}\BibitemShut
  {NoStop}%
\bibitem [{\citenamefont {Nakatsuji}\ \emph {et~al.}(2015)\citenamefont
  {Nakatsuji}, \citenamefont {Kiyohara},\ and\ \citenamefont
  {Higo}}]{Nakatsuji2015}%
  \BibitemOpen
  \bibfield  {author} {\bibinfo {author} {\bibfnamefont {S.}~\bibnamefont
  {Nakatsuji}}, \bibinfo {author} {\bibfnamefont {N.}~\bibnamefont
  {Kiyohara}},\ and\ \bibinfo {author} {\bibfnamefont {T.}~\bibnamefont
  {Higo}},\ }\bibfield  {title} {\bibinfo {title} {{Large anomalous Hall effect
  in a non-collinear antiferromagnet at room temperature}},\ }\href
  {https://doi.org/10.1038/nature15723} {\bibfield  {journal} {\bibinfo
  {journal} {Nature}\ }\textbf {\bibinfo {volume} {527}},\ \bibinfo {pages}
  {212} (\bibinfo {year} {2015})}\BibitemShut {NoStop}%
\bibitem [{\citenamefont {Grohol}\ \emph {et~al.}(2005)\citenamefont {Grohol},
  \citenamefont {Matan}, \citenamefont {Cho}, \citenamefont {Lee},
  \citenamefont {Lynn}, \citenamefont {Nocera},\ and\ \citenamefont
  {Lee}}]{Grohol2005}%
  \BibitemOpen
  \bibfield  {author} {\bibinfo {author} {\bibfnamefont {D.}~\bibnamefont
  {Grohol}}, \bibinfo {author} {\bibfnamefont {K.}~\bibnamefont {Matan}},
  \bibinfo {author} {\bibfnamefont {J.-H.}\ \bibnamefont {Cho}}, \bibinfo
  {author} {\bibfnamefont {S.-H.}\ \bibnamefont {Lee}}, \bibinfo {author}
  {\bibfnamefont {J.~W.}\ \bibnamefont {Lynn}}, \bibinfo {author}
  {\bibfnamefont {D.~G.}\ \bibnamefont {Nocera}},\ and\ \bibinfo {author}
  {\bibfnamefont {Y.~S.}\ \bibnamefont {Lee}},\ }\bibfield  {title} {\bibinfo
  {title} {Spin chirality on a two-dimensional frustrated lattice},\ }\href
  {https://doi.org/10.1038/nmat1353} {\bibfield  {journal} {\bibinfo  {journal}
  {Nat. Mater.}\ }\textbf {\bibinfo {volume} {4}},\ \bibinfo {pages} {323}
  (\bibinfo {year} {2005})}\BibitemShut {NoStop}%
\bibitem [{\citenamefont {Xu}\ \emph {et~al.}(2015)\citenamefont {Xu},
  \citenamefont {Lian},\ and\ \citenamefont {Zhang}}]{Xu2015}%
  \BibitemOpen
  \bibfield  {author} {\bibinfo {author} {\bibfnamefont {G.}~\bibnamefont
  {Xu}}, \bibinfo {author} {\bibfnamefont {B.}~\bibnamefont {Lian}},\ and\
  \bibinfo {author} {\bibfnamefont {S.-C.}\ \bibnamefont {Zhang}},\ }\bibfield
  {title} {\bibinfo {title} {Intrinsic quantum anomalous {H}all effect in the
  kagome lattice {Cs}$_2${LiMn}$_3${F}$_{12}$},\ }\href
  {https://doi.org/10.1103/PhysRevLett.115.186802} {\bibfield  {journal}
  {\bibinfo  {journal} {Phys. Rev. Lett.}\ }\textbf {\bibinfo {volume} {115}},\
  \bibinfo {pages} {186802} (\bibinfo {year} {2015})}\BibitemShut {NoStop}%
\bibitem [{\citenamefont {Schlickum}\ \emph {et~al.}(2008)\citenamefont
  {Schlickum}, \citenamefont {Decker}, \citenamefont {Klappenberger},
  \citenamefont {Zoppellaro}, \citenamefont {Klyatskaya}, \citenamefont
  {Auw\"arter}, \citenamefont {Neppl}, \citenamefont {Kern}, \citenamefont
  {Brune}, \citenamefont {Ruben},\ and\ \citenamefont {Barth}}]{Schlickum2008}%
  \BibitemOpen
  \bibfield  {author} {\bibinfo {author} {\bibfnamefont {U.}~\bibnamefont
  {Schlickum}}, \bibinfo {author} {\bibfnamefont {R.}~\bibnamefont {Decker}},
  \bibinfo {author} {\bibfnamefont {F.}~\bibnamefont {Klappenberger}}, \bibinfo
  {author} {\bibfnamefont {G.}~\bibnamefont {Zoppellaro}}, \bibinfo {author}
  {\bibfnamefont {S.}~\bibnamefont {Klyatskaya}}, \bibinfo {author}
  {\bibfnamefont {W.}~\bibnamefont {Auw\"arter}}, \bibinfo {author}
  {\bibfnamefont {S.}~\bibnamefont {Neppl}}, \bibinfo {author} {\bibfnamefont
  {K.}~\bibnamefont {Kern}}, \bibinfo {author} {\bibfnamefont {H.}~\bibnamefont
  {Brune}}, \bibinfo {author} {\bibfnamefont {M.}~\bibnamefont {Ruben}},\ and\
  \bibinfo {author} {\bibfnamefont {J.~V.}\ \bibnamefont {Barth}},\ }\bibfield
  {title} {\bibinfo {title} {Chiral kagome lattice from simple ditopic
  molecular bricks},\ }\href {https://doi.org/10.1021/ja8028119} {\bibfield
  {journal} {\bibinfo  {journal} {J. Am. Chem. Soc.}\ }\textbf {\bibinfo
  {volume} {130}},\ \bibinfo {pages} {11778} (\bibinfo {year}
  {2008})}\BibitemShut {NoStop}%
\bibitem [{\citenamefont {Shi}\ and\ \citenamefont {Lin}(2009)}]{Shi2009}%
  \BibitemOpen
  \bibfield  {author} {\bibinfo {author} {\bibfnamefont {Z.}~\bibnamefont
  {Shi}}\ and\ \bibinfo {author} {\bibfnamefont {N.}~\bibnamefont {Lin}},\
  }\bibfield  {title} {\bibinfo {title} {Porphyrin-based two-dimensional
  coordination kagome lattice self-assembled on a {Au}(111) surface},\ }\href
  {https://doi.org/10.1021/ja900499b} {\bibfield  {journal} {\bibinfo
  {journal} {J. Am. Chem. Soc.}\ }\textbf {\bibinfo {volume} {131}},\ \bibinfo
  {pages} {5376} (\bibinfo {year} {2009})}\BibitemShut {NoStop}%
\bibitem [{\citenamefont {Yan}\ \emph {et~al.}(2021{\natexlab{a}})\citenamefont
  {Yan}, \citenamefont {Silveira}, \citenamefont {Alldritt}, \citenamefont
  {Krej\v{c}\'i}, \citenamefont {Foster},\ and\ \citenamefont
  {Liljeroth}}]{Yan2021synthesis}%
  \BibitemOpen
  \bibfield  {author} {\bibinfo {author} {\bibfnamefont {L.}~\bibnamefont
  {Yan}}, \bibinfo {author} {\bibfnamefont {O.~J.}\ \bibnamefont {Silveira}},
  \bibinfo {author} {\bibfnamefont {B.}~\bibnamefont {Alldritt}}, \bibinfo
  {author} {\bibfnamefont {O.}~\bibnamefont {Krej\v{c}\'i}}, \bibinfo {author}
  {\bibfnamefont {A.~S.}\ \bibnamefont {Foster}},\ and\ \bibinfo {author}
  {\bibfnamefont {P.}~\bibnamefont {Liljeroth}},\ }\bibfield  {title} {\bibinfo
  {title} {Synthesis and local probe gating of a monolayer metal-organic
  framework},\ }\href {https://doi.org/https://doi.org/10.1002/adfm.202100519}
  {\bibfield  {journal} {\bibinfo  {journal} {Adv. Funct. Mater.}\ }\textbf
  {\bibinfo {volume} {31}},\ \bibinfo {pages} {2100519} (\bibinfo {year}
  {2021}{\natexlab{a}})}\BibitemShut {NoStop}%
\bibitem [{\citenamefont {Kumar}\ \emph {et~al.}(2021)\citenamefont {Kumar},
  \citenamefont {Hellerstedt}, \citenamefont {Field}, \citenamefont {Lowe},
  \citenamefont {Yin}, \citenamefont {Medhekar},\ and\ \citenamefont
  {Schiffrin}}]{Kumar2021}%
  \BibitemOpen
  \bibfield  {author} {\bibinfo {author} {\bibfnamefont {D.}~\bibnamefont
  {Kumar}}, \bibinfo {author} {\bibfnamefont {J.}~\bibnamefont {Hellerstedt}},
  \bibinfo {author} {\bibfnamefont {B.}~\bibnamefont {Field}}, \bibinfo
  {author} {\bibfnamefont {B.}~\bibnamefont {Lowe}}, \bibinfo {author}
  {\bibfnamefont {Y.}~\bibnamefont {Yin}}, \bibinfo {author} {\bibfnamefont
  {N.~V.}\ \bibnamefont {Medhekar}},\ and\ \bibinfo {author} {\bibfnamefont
  {A.}~\bibnamefont {Schiffrin}},\ }\bibfield  {title} {\bibinfo {title}
  {Manifestation of strongly correlated electrons in a {2D} kagome
  metal{\textendash}organic framework},\ }\href
  {https://doi.org/10.1002/adfm.202106474} {\bibfield  {journal} {\bibinfo
  {journal} {Adv. Funct. Mater.}\ }\textbf {\bibinfo {volume} {31}},\ \bibinfo
  {pages} {2106474} (\bibinfo {year} {2021})}\BibitemShut {NoStop}%
\bibitem [{\citenamefont {Field}\ \emph {et~al.}(2022)\citenamefont {Field},
  \citenamefont {Schiffrin},\ and\ \citenamefont {Medhekar}}]{Field2022}%
  \BibitemOpen
  \bibfield  {author} {\bibinfo {author} {\bibfnamefont {B.}~\bibnamefont
  {Field}}, \bibinfo {author} {\bibfnamefont {A.}~\bibnamefont {Schiffrin}},\
  and\ \bibinfo {author} {\bibfnamefont {N.~V.}\ \bibnamefont {Medhekar}},\
  }\bibfield  {title} {\bibinfo {title} {Correlation-induced magnetism in
  substrate-supported 2{D} metal-organic frameworks},\ }\href
  {https://doi.org/10.1038/s41524-022-00918-0} {\bibfield  {journal} {\bibinfo
  {journal} {Npj Comput. Mater.}\ }\textbf {\bibinfo {volume} {8}},\ \bibinfo
  {pages} {227} (\bibinfo {year} {2022})}\BibitemShut {NoStop}%
\bibitem [{\citenamefont {Yu}(1965)}]{Yu1965}%
  \BibitemOpen
  \bibfield  {author} {\bibinfo {author} {\bibfnamefont {L.}~\bibnamefont
  {Yu}},\ }\bibfield  {title} {\bibinfo {title} {Bound state in superconductors
  with paramagnetic impurities},\ }\href@noop {} {\bibfield  {journal}
  {\bibinfo  {journal} {Acta Phys. Sin.}\ }\textbf {\bibinfo {volume} {21}},\
  \bibinfo {pages} {75} (\bibinfo {year} {1965})}\BibitemShut {NoStop}%
\bibitem [{\citenamefont {Shiba}(1968)}]{Shiba1968}%
  \BibitemOpen
  \bibfield  {author} {\bibinfo {author} {\bibfnamefont {H.}~\bibnamefont
  {Shiba}},\ }\bibfield  {title} {\bibinfo {title} {Classical spins in
  superconductors},\ }\href {https://doi.org/10.1143/PTP.40.435} {\bibfield
  {journal} {\bibinfo  {journal} {Prog. of Theo. Phys.}\ }\textbf {\bibinfo
  {volume} {40}},\ \bibinfo {pages} {435} (\bibinfo {year} {1968})}\BibitemShut
  {NoStop}%
\bibitem [{\citenamefont {Rusinov}(1969)}]{Rusinov1969}%
  \BibitemOpen
  \bibfield  {author} {\bibinfo {author} {\bibfnamefont {A.~I.}\ \bibnamefont
  {Rusinov}},\ }\bibfield  {title} {\bibinfo {title} {Superconductivity near a
  paramagnetic impurity},\ }\href
  {http://www.jetpletters.ac.ru/ps/1658/article{\_}25295.shtml} {\bibfield
  {journal} {\bibinfo  {journal} {JETP Lett.}\ }\textbf {\bibinfo {volume}
  {9}},\ \bibinfo {pages} {85} (\bibinfo {year} {1969})}\BibitemShut {NoStop}%
\bibitem [{\citenamefont {Yao}\ \emph {et~al.}(2014)\citenamefont {Yao},
  \citenamefont {Moca}, \citenamefont {Weymann}, \citenamefont {Sau},
  \citenamefont {Lukin}, \citenamefont {Demler},\ and\ \citenamefont
  {Zar\'and}}]{Yao2014a}%
  \BibitemOpen
  \bibfield  {author} {\bibinfo {author} {\bibfnamefont {N.~Y.}\ \bibnamefont
  {Yao}}, \bibinfo {author} {\bibfnamefont {C.~P.}\ \bibnamefont {Moca}},
  \bibinfo {author} {\bibfnamefont {I.}~\bibnamefont {Weymann}}, \bibinfo
  {author} {\bibfnamefont {J.~D.}\ \bibnamefont {Sau}}, \bibinfo {author}
  {\bibfnamefont {M.~D.}\ \bibnamefont {Lukin}}, \bibinfo {author}
  {\bibfnamefont {E.~A.}\ \bibnamefont {Demler}},\ and\ \bibinfo {author}
  {\bibfnamefont {G.}~\bibnamefont {Zar\'and}},\ }\bibfield  {title} {\bibinfo
  {title} {{Phase diagram and excitations of a Shiba molecule}},\ }\href
  {https://doi.org/10.1103/PhysRevB.90.241108} {\bibfield  {journal} {\bibinfo
  {journal} {Phys. Rev. B}\ }\textbf {\bibinfo {volume} {90}},\ \bibinfo
  {pages} {241108} (\bibinfo {year} {2014})}\BibitemShut {NoStop}%
\bibitem [{\citenamefont {Ruby}\ \emph {et~al.}(2018)\citenamefont {Ruby},
  \citenamefont {Heinrich}, \citenamefont {Peng}, \citenamefont {von Oppen},\
  and\ \citenamefont {Franke}}]{Ruby2018}%
  \BibitemOpen
  \bibfield  {author} {\bibinfo {author} {\bibfnamefont {M.}~\bibnamefont
  {Ruby}}, \bibinfo {author} {\bibfnamefont {B.~W.}\ \bibnamefont {Heinrich}},
  \bibinfo {author} {\bibfnamefont {Y.}~\bibnamefont {Peng}}, \bibinfo {author}
  {\bibfnamefont {F.}~\bibnamefont {von Oppen}},\ and\ \bibinfo {author}
  {\bibfnamefont {K.~J.}\ \bibnamefont {Franke}},\ }\bibfield  {title}
  {\bibinfo {title} {Wave-function hybridization in {Y}u-{S}hiba-{R}usinov
  dimers},\ }\href {https://doi.org/10.1103/PhysRevLett.120.156803} {\bibfield
  {journal} {\bibinfo  {journal} {Phys. Rev. Lett.}\ }\textbf {\bibinfo
  {volume} {120}},\ \bibinfo {pages} {156803} (\bibinfo {year}
  {2018})}\BibitemShut {NoStop}%
\bibitem [{\citenamefont {Choi}\ \emph {et~al.}(2018)\citenamefont {Choi},
  \citenamefont {Fern\'andez}, \citenamefont {Herrera}, \citenamefont
  {Rubio-Verd\'u}, \citenamefont {Ugeda}, \citenamefont {Guillam\'on},
  \citenamefont {Suderow}, \citenamefont {Pascual},\ and\ \citenamefont
  {Lorente}}]{Choi2018}%
  \BibitemOpen
  \bibfield  {author} {\bibinfo {author} {\bibfnamefont {D.-J.}\ \bibnamefont
  {Choi}}, \bibinfo {author} {\bibfnamefont {C.~G.}\ \bibnamefont
  {Fern\'andez}}, \bibinfo {author} {\bibfnamefont {E.}~\bibnamefont
  {Herrera}}, \bibinfo {author} {\bibfnamefont {C.}~\bibnamefont
  {Rubio-Verd\'u}}, \bibinfo {author} {\bibfnamefont {M.~M.}\ \bibnamefont
  {Ugeda}}, \bibinfo {author} {\bibfnamefont {I.}~\bibnamefont {Guillam\'on}},
  \bibinfo {author} {\bibfnamefont {H.}~\bibnamefont {Suderow}}, \bibinfo
  {author} {\bibfnamefont {J.~I.}\ \bibnamefont {Pascual}},\ and\ \bibinfo
  {author} {\bibfnamefont {N.}~\bibnamefont {Lorente}},\ }\bibfield  {title}
  {\bibinfo {title} {Influence of magnetic ordering between {C}r adatoms on the
  {Y}u-{S}hiba-{R}usinov states of the
  $\ensuremath{\beta}\text{\ensuremath{-}}\mathrm{Bi}_{2}\mathrm{Pd}$
  superconductor},\ }\href {https://doi.org/10.1103/PhysRevLett.120.167001}
  {\bibfield  {journal} {\bibinfo  {journal} {Phys. Rev. Lett.}\ }\textbf
  {\bibinfo {volume} {120}},\ \bibinfo {pages} {167001} (\bibinfo {year}
  {2018})}\BibitemShut {NoStop}%
\bibitem [{\citenamefont {Kim}\ \emph {et~al.}(2018)\citenamefont {Kim},
  \citenamefont {Palacio-Morales}, \citenamefont {Posske}, \citenamefont
  {Rozsa}, \citenamefont {Palotas}, \citenamefont {Szunyogh}, \citenamefont
  {Thorwart},\ and\ \citenamefont {Wiesendanger}}]{Kim2018}%
  \BibitemOpen
  \bibfield  {author} {\bibinfo {author} {\bibfnamefont {H.}~\bibnamefont
  {Kim}}, \bibinfo {author} {\bibfnamefont {A.}~\bibnamefont
  {Palacio-Morales}}, \bibinfo {author} {\bibfnamefont {T.}~\bibnamefont
  {Posske}}, \bibinfo {author} {\bibfnamefont {L.}~\bibnamefont {Rozsa}},
  \bibinfo {author} {\bibfnamefont {K.}~\bibnamefont {Palotas}}, \bibinfo
  {author} {\bibfnamefont {L.}~\bibnamefont {Szunyogh}}, \bibinfo {author}
  {\bibfnamefont {M.}~\bibnamefont {Thorwart}},\ and\ \bibinfo {author}
  {\bibfnamefont {R.}~\bibnamefont {Wiesendanger}},\ }\bibfield  {title}
  {\bibinfo {title} {Toward tailoring {M}ajorana bound states in artificially
  constructed magnetic atom chains on elemental superconductors},\ }\href
  {https://doi.org/10.1126/sciadv.aar5251} {\bibfield  {journal} {\bibinfo
  {journal} {Sci. Adv.}\ }\textbf {\bibinfo {volume} {4}},\ \bibinfo {pages}
  {eaar5251} (\bibinfo {year} {2018})}\BibitemShut {NoStop}%
\bibitem [{\citenamefont {Kezilebieke}\ \emph {et~al.}(2018)\citenamefont
  {Kezilebieke}, \citenamefont {Dvorak}, \citenamefont {Ojanen},\ and\
  \citenamefont {Liljeroth}}]{Kezilebieke2018}%
  \BibitemOpen
  \bibfield  {author} {\bibinfo {author} {\bibfnamefont {S.}~\bibnamefont
  {Kezilebieke}}, \bibinfo {author} {\bibfnamefont {M.}~\bibnamefont {Dvorak}},
  \bibinfo {author} {\bibfnamefont {T.}~\bibnamefont {Ojanen}},\ and\ \bibinfo
  {author} {\bibfnamefont {P.}~\bibnamefont {Liljeroth}},\ }\bibfield  {title}
  {\bibinfo {title} {{Coupled Yu-Shiba-Rusinov states in molecular dimers on
  {NbSe}$_2$}},\ }\href {https://doi.org/10.1021/acs.nanolett.7b05050}
  {\bibfield  {journal} {\bibinfo  {journal} {Nano Lett.}\ }\textbf {\bibinfo
  {volume} {18}},\ \bibinfo {pages} {2311} (\bibinfo {year}
  {2018})}\BibitemShut {NoStop}%
\bibitem [{\citenamefont {Pientka}\ \emph {et~al.}(2013)\citenamefont
  {Pientka}, \citenamefont {Glazman},\ and\ \citenamefont {von
  Oppen}}]{Pientka2013}%
  \BibitemOpen
  \bibfield  {author} {\bibinfo {author} {\bibfnamefont {F.}~\bibnamefont
  {Pientka}}, \bibinfo {author} {\bibfnamefont {L.~I.}\ \bibnamefont
  {Glazman}},\ and\ \bibinfo {author} {\bibfnamefont {F.}~\bibnamefont {von
  Oppen}},\ }\bibfield  {title} {\bibinfo {title} {{Topological superconducting
  phase in helical Shiba chains}},\ }\href
  {https://doi.org/10.1103/PhysRevB.88.155420} {\bibfield  {journal} {\bibinfo
  {journal} {Phys. Rev. B}\ }\textbf {\bibinfo {volume} {88}},\ \bibinfo
  {pages} {155420} (\bibinfo {year} {2013})}\BibitemShut {NoStop}%
\bibitem [{\citenamefont {Liebhaber}\ \emph {et~al.}(2022)\citenamefont
  {Liebhaber}, \citenamefont {R{\"u}tten}, \citenamefont {Reecht},
  \citenamefont {Steiner}, \citenamefont {Rohlf}, \citenamefont {Rossnagel},
  \citenamefont {von Oppen},\ and\ \citenamefont {Franke}}]{Liebhaber2022}%
  \BibitemOpen
  \bibfield  {author} {\bibinfo {author} {\bibfnamefont {E.}~\bibnamefont
  {Liebhaber}}, \bibinfo {author} {\bibfnamefont {L.~M.}\ \bibnamefont
  {R{\"u}tten}}, \bibinfo {author} {\bibfnamefont {G.}~\bibnamefont {Reecht}},
  \bibinfo {author} {\bibfnamefont {J.~F.}\ \bibnamefont {Steiner}}, \bibinfo
  {author} {\bibfnamefont {S.}~\bibnamefont {Rohlf}}, \bibinfo {author}
  {\bibfnamefont {K.}~\bibnamefont {Rossnagel}}, \bibinfo {author}
  {\bibfnamefont {F.}~\bibnamefont {von Oppen}},\ and\ \bibinfo {author}
  {\bibfnamefont {K.~J.}\ \bibnamefont {Franke}},\ }\bibfield  {title}
  {\bibinfo {title} {Quantum spins and hybridization in
  artificially-constructed chains of magnetic adatoms on a superconductor},\
  }\href {https://doi.org/10.1038/s41467-022-29879-0} {\bibfield  {journal}
  {\bibinfo  {journal} {Nat. Commun.}\ }\textbf {\bibinfo {volume} {13}},\
  \bibinfo {pages} {1} (\bibinfo {year} {2022})}\BibitemShut {NoStop}%
\bibitem [{\citenamefont {Schneider}\ \emph {et~al.}(2022)\citenamefont
  {Schneider}, \citenamefont {Beck}, \citenamefont {Neuhaus-Steinmetz},
  \citenamefont {R{\'o}zsa}, \citenamefont {Posske}, \citenamefont {Wiebe},\
  and\ \citenamefont {Wiesendanger}}]{Schneider2022}%
  \BibitemOpen
  \bibfield  {author} {\bibinfo {author} {\bibfnamefont {L.}~\bibnamefont
  {Schneider}}, \bibinfo {author} {\bibfnamefont {P.}~\bibnamefont {Beck}},
  \bibinfo {author} {\bibfnamefont {J.}~\bibnamefont {Neuhaus-Steinmetz}},
  \bibinfo {author} {\bibfnamefont {L.}~\bibnamefont {R{\'o}zsa}}, \bibinfo
  {author} {\bibfnamefont {T.}~\bibnamefont {Posske}}, \bibinfo {author}
  {\bibfnamefont {J.}~\bibnamefont {Wiebe}},\ and\ \bibinfo {author}
  {\bibfnamefont {R.}~\bibnamefont {Wiesendanger}},\ }\bibfield  {title}
  {\bibinfo {title} {Precursors of {M}ajorana modes and their length-dependent
  energy oscillations probed at both ends of atomic {S}hiba chains},\ }\href
  {https://doi.org/10.1038/s41565-022-01078-4} {\bibfield  {journal} {\bibinfo
  {journal} {Nat. Nanotechnol.}\ }\textbf {\bibinfo {volume} {17}},\ \bibinfo
  {pages} {384} (\bibinfo {year} {2022})}\BibitemShut {NoStop}%
\bibitem [{\citenamefont {Steiner}\ \emph {et~al.}(2022)\citenamefont
  {Steiner}, \citenamefont {Mora}, \citenamefont {Franke},\ and\ \citenamefont
  {von Oppen}}]{Steiner2021}%
  \BibitemOpen
  \bibfield  {author} {\bibinfo {author} {\bibfnamefont {J.~F.}\ \bibnamefont
  {Steiner}}, \bibinfo {author} {\bibfnamefont {C.}~\bibnamefont {Mora}},
  \bibinfo {author} {\bibfnamefont {K.~J.}\ \bibnamefont {Franke}},\ and\
  \bibinfo {author} {\bibfnamefont {F.}~\bibnamefont {von Oppen}},\ }\bibfield
  {title} {\bibinfo {title} {Quantum magnetism and topological
  superconductivity in {Y}u-{S}hiba-{R}usinov chains},\ }\href
  {https://doi.org/10.1103/PhysRevLett.128.036801} {\bibfield  {journal}
  {\bibinfo  {journal} {Phys. Rev. Lett.}\ }\textbf {\bibinfo {volume} {128}},\
  \bibinfo {pages} {036801} (\bibinfo {year} {2022})}\BibitemShut {NoStop}%
\bibitem [{\citenamefont {Lin}\ \emph {et~al.}(2022)\citenamefont {Lin},
  \citenamefont {Chen}, \citenamefont {Kumar}, \citenamefont {Yeh},
  \citenamefont {Lin}, \citenamefont {Bl\"ugel}, \citenamefont {Bihlmayer},\
  and\ \citenamefont {Hsu}}]{Lin2022}%
  \BibitemOpen
  \bibfield  {author} {\bibinfo {author} {\bibfnamefont {Y.-H.}\ \bibnamefont
  {Lin}}, \bibinfo {author} {\bibfnamefont {C.-J.}\ \bibnamefont {Chen}},
  \bibinfo {author} {\bibfnamefont {N.}~\bibnamefont {Kumar}}, \bibinfo
  {author} {\bibfnamefont {T.-Y.}\ \bibnamefont {Yeh}}, \bibinfo {author}
  {\bibfnamefont {T.-H.}\ \bibnamefont {Lin}}, \bibinfo {author} {\bibfnamefont
  {S.}~\bibnamefont {Bl\"ugel}}, \bibinfo {author} {\bibfnamefont
  {G.}~\bibnamefont {Bihlmayer}},\ and\ \bibinfo {author} {\bibfnamefont
  {P.-J.}\ \bibnamefont {Hsu}},\ }\bibfield  {title} {\bibinfo {title}
  {Fabrication and imaging monatomic {Ni} kagome lattice on superconducting
  {Pb}(111)},\ }\href {https://doi.org/10.1021/acs.nanolett.2c02831} {\bibfield
   {journal} {\bibinfo  {journal} {Nano Lett.}\ }\textbf {\bibinfo {volume}
  {22}},\ \bibinfo {pages} {8475} (\bibinfo {year} {2022})}\BibitemShut
  {NoStop}%
\bibitem [{\citenamefont {Yan}\ \emph {et~al.}(2021{\natexlab{b}})\citenamefont
  {Yan}, \citenamefont {Silveira}, \citenamefont {Alldritt}, \citenamefont
  {Kezilebieke}, \citenamefont {Foster},\ and\ \citenamefont
  {Liljeroth}}]{Yan2021}%
  \BibitemOpen
  \bibfield  {author} {\bibinfo {author} {\bibfnamefont {L.}~\bibnamefont
  {Yan}}, \bibinfo {author} {\bibfnamefont {O.~J.}\ \bibnamefont {Silveira}},
  \bibinfo {author} {\bibfnamefont {B.}~\bibnamefont {Alldritt}}, \bibinfo
  {author} {\bibfnamefont {S.}~\bibnamefont {Kezilebieke}}, \bibinfo {author}
  {\bibfnamefont {A.~S.}\ \bibnamefont {Foster}},\ and\ \bibinfo {author}
  {\bibfnamefont {P.}~\bibnamefont {Liljeroth}},\ }\bibfield  {title} {\bibinfo
  {title} {Two-dimensional metal-organic framework on superconducting
  {NbSe}$_2$},\ }\href {https://doi.org/10.1021/acsnano.1c05986} {\bibfield
  {journal} {\bibinfo  {journal} {{ACS} Nano}\ }\textbf {\bibinfo {volume}
  {15}},\ \bibinfo {pages} {17813} (\bibinfo {year}
  {2021}{\natexlab{b}})}\BibitemShut {NoStop}%
\bibitem [{\citenamefont {Franke}\ \emph {et~al.}(2011)\citenamefont {Franke},
  \citenamefont {Schulze},\ and\ \citenamefont {Pascual}}]{Franke2011}%
  \BibitemOpen
  \bibfield  {author} {\bibinfo {author} {\bibfnamefont {K.~J.}\ \bibnamefont
  {Franke}}, \bibinfo {author} {\bibfnamefont {G.}~\bibnamefont {Schulze}},\
  and\ \bibinfo {author} {\bibfnamefont {J.~I.}\ \bibnamefont {Pascual}},\
  }\bibfield  {title} {\bibinfo {title} {{Competition of superconducting
  phenomena and Kondo screening at the nanoscale}},\ }\href
  {https://doi.org/10.1126/science.1202204} {\bibfield  {journal} {\bibinfo
  {journal} {Science}\ }\textbf {\bibinfo {volume} {332}},\ \bibinfo {pages}
  {940} (\bibinfo {year} {2011})}\BibitemShut {NoStop}%
\bibitem [{\citenamefont {Farinacci}\ \emph {et~al.}(2018)\citenamefont
  {Farinacci}, \citenamefont {Ahmadi}, \citenamefont {Reecht}, \citenamefont
  {Ruby}, \citenamefont {Bogdanoff}, \citenamefont {Peters}, \citenamefont
  {Heinrich}, \citenamefont {von Oppen},\ and\ \citenamefont
  {Franke}}]{Farinacci2018}%
  \BibitemOpen
  \bibfield  {author} {\bibinfo {author} {\bibfnamefont {L.}~\bibnamefont
  {Farinacci}}, \bibinfo {author} {\bibfnamefont {G.}~\bibnamefont {Ahmadi}},
  \bibinfo {author} {\bibfnamefont {G.}~\bibnamefont {Reecht}}, \bibinfo
  {author} {\bibfnamefont {M.}~\bibnamefont {Ruby}}, \bibinfo {author}
  {\bibfnamefont {N.}~\bibnamefont {Bogdanoff}}, \bibinfo {author}
  {\bibfnamefont {O.}~\bibnamefont {Peters}}, \bibinfo {author} {\bibfnamefont
  {B.~W.}\ \bibnamefont {Heinrich}}, \bibinfo {author} {\bibfnamefont
  {F.}~\bibnamefont {von Oppen}},\ and\ \bibinfo {author} {\bibfnamefont
  {K.~J.}\ \bibnamefont {Franke}},\ }\bibfield  {title} {\bibinfo {title}
  {{Tuning the coupling of an individual magnetic impurity to a superconductor:
  Quantum phase transition and transport}},\ }\href
  {https://doi.org/10.1103/PhysRevLett.121.196803} {\bibfield  {journal}
  {\bibinfo  {journal} {Phys. Rev. Lett.}\ }\textbf {\bibinfo {volume} {121}},\
  \bibinfo {pages} {196803} (\bibinfo {year} {2018})}\BibitemShut {NoStop}%
\bibitem [{\citenamefont {Homberg}\ \emph {et~al.}(2020)\citenamefont
  {Homberg}, \citenamefont {Weismann}, \citenamefont {Berndt},\ and\
  \citenamefont {Gruber}}]{Homberg2020}%
  \BibitemOpen
  \bibfield  {author} {\bibinfo {author} {\bibfnamefont {J.}~\bibnamefont
  {Homberg}}, \bibinfo {author} {\bibfnamefont {A.}~\bibnamefont {Weismann}},
  \bibinfo {author} {\bibfnamefont {R.}~\bibnamefont {Berndt}},\ and\ \bibinfo
  {author} {\bibfnamefont {M.}~\bibnamefont {Gruber}},\ }\bibfield  {title}
  {\bibinfo {title} {Inducing and controlling molecular magnetism through
  supramolecular manipulation},\ }\href
  {https://doi.org/10.1021/acsnano.0c07574} {\bibfield  {journal} {\bibinfo
  {journal} {ACS Nano}\ }\textbf {\bibinfo {volume} {14}},\ \bibinfo {pages}
  {17387} (\bibinfo {year} {2020})}\BibitemShut {NoStop}%
\bibitem [{\citenamefont {Lu}\ \emph {et~al.}(2021)\citenamefont {Lu},
  \citenamefont {Nam}, \citenamefont {Xiao}, \citenamefont {Liu}, \citenamefont
  {Guo}, \citenamefont {Bai}, \citenamefont {Cheng}, \citenamefont {Deng},
  \citenamefont {Li}, \citenamefont {Zhou}, \citenamefont {Henkelman},
  \citenamefont {Fiete}, \citenamefont {Gao}, \citenamefont {MacDonald},
  \citenamefont {Zhang},\ and\ \citenamefont {Shih}}]{Lu2021}%
  \BibitemOpen
  \bibfield  {author} {\bibinfo {author} {\bibfnamefont {S.}~\bibnamefont
  {Lu}}, \bibinfo {author} {\bibfnamefont {H.}~\bibnamefont {Nam}}, \bibinfo
  {author} {\bibfnamefont {P.}~\bibnamefont {Xiao}}, \bibinfo {author}
  {\bibfnamefont {M.}~\bibnamefont {Liu}}, \bibinfo {author} {\bibfnamefont
  {Y.}~\bibnamefont {Guo}}, \bibinfo {author} {\bibfnamefont {Y.}~\bibnamefont
  {Bai}}, \bibinfo {author} {\bibfnamefont {Z.}~\bibnamefont {Cheng}}, \bibinfo
  {author} {\bibfnamefont {J.}~\bibnamefont {Deng}}, \bibinfo {author}
  {\bibfnamefont {Y.}~\bibnamefont {Li}}, \bibinfo {author} {\bibfnamefont
  {H.}~\bibnamefont {Zhou}}, \bibinfo {author} {\bibfnamefont {G.}~\bibnamefont
  {Henkelman}}, \bibinfo {author} {\bibfnamefont {G.~A.}\ \bibnamefont
  {Fiete}}, \bibinfo {author} {\bibfnamefont {H.-J.}\ \bibnamefont {Gao}},
  \bibinfo {author} {\bibfnamefont {A.~H.}\ \bibnamefont {MacDonald}}, \bibinfo
  {author} {\bibfnamefont {C.}~\bibnamefont {Zhang}},\ and\ \bibinfo {author}
  {\bibfnamefont {C.-K.}\ \bibnamefont {Shih}},\ }\bibfield  {title} {\bibinfo
  {title} {{PTCDA} molecular monolayer on {Pb} thin films: An unusual
  $\ensuremath{\pi}$-electron {K}ondo system and its interplay with a
  quantum-confined superconductor},\ }\href
  {https://doi.org/10.1103/PhysRevLett.127.186805} {\bibfield  {journal}
  {\bibinfo  {journal} {Phys. Rev. Lett.}\ }\textbf {\bibinfo {volume} {127}},\
  \bibinfo {pages} {186805} (\bibinfo {year} {2021})}\BibitemShut {NoStop}%
\bibitem [{\citenamefont {Frank}\ and\ \citenamefont
  {Jacob}(2015)}]{Frank2015}%
  \BibitemOpen
  \bibfield  {author} {\bibinfo {author} {\bibfnamefont {S.}~\bibnamefont
  {Frank}}\ and\ \bibinfo {author} {\bibfnamefont {D.}~\bibnamefont {Jacob}},\
  }\bibfield  {title} {\bibinfo {title} {Orbital signatures of {Fano-Kondo}
  line shapes in {STM} adatom spectroscopy},\ }\href
  {https://doi.org/10.1103/PhysRevB.92.235127} {\bibfield  {journal} {\bibinfo
  {journal} {Phys. Rev. B}\ }\textbf {\bibinfo {volume} {92}},\ \bibinfo
  {pages} {235127} (\bibinfo {year} {2015})}\BibitemShut {NoStop}%
\bibitem [{\citenamefont {Farinacci}\ \emph {et~al.}(2020)\citenamefont
  {Farinacci}, \citenamefont {Ahmadi}, \citenamefont {Ruby}, \citenamefont
  {Reecht}, \citenamefont {Heinrich}, \citenamefont {Czekelius}, \citenamefont
  {von Oppen},\ and\ \citenamefont {Franke}}]{Farinacci2020}%
  \BibitemOpen
  \bibfield  {author} {\bibinfo {author} {\bibfnamefont {L.}~\bibnamefont
  {Farinacci}}, \bibinfo {author} {\bibfnamefont {G.}~\bibnamefont {Ahmadi}},
  \bibinfo {author} {\bibfnamefont {M.}~\bibnamefont {Ruby}}, \bibinfo {author}
  {\bibfnamefont {G.}~\bibnamefont {Reecht}}, \bibinfo {author} {\bibfnamefont
  {B.~W.}\ \bibnamefont {Heinrich}}, \bibinfo {author} {\bibfnamefont
  {C.}~\bibnamefont {Czekelius}}, \bibinfo {author} {\bibfnamefont
  {F.}~\bibnamefont {von Oppen}},\ and\ \bibinfo {author} {\bibfnamefont
  {K.~J.}\ \bibnamefont {Franke}},\ }\bibfield  {title} {\bibinfo {title}
  {Interfering tunneling paths through magnetic molecules on superconductors:
  Asymmetries of {K}ondo and {Y}u-{S}hiba-{R}usinov resonances},\ }\href
  {https://doi.org/10.1103/PhysRevLett.121.196803} {\bibfield  {journal}
  {\bibinfo  {journal} {Phys. Rev. Lett.}\ }\textbf {\bibinfo {volume} {125}},\
  \bibinfo {pages} {256805} (\bibinfo {year} {2020})}\BibitemShut {NoStop}%
\bibitem [{\citenamefont {Liebhaber}\ \emph {et~al.}(2019)\citenamefont
  {Liebhaber}, \citenamefont {Acero~Gonzalez}, \citenamefont {Baba},
  \citenamefont {Reecht}, \citenamefont {Heinrich}, \citenamefont {Rohlf},
  \citenamefont {Rossnagel}, \citenamefont {von Oppen},\ and\ \citenamefont
  {Franke}}]{Liebhaber2019}%
  \BibitemOpen
  \bibfield  {author} {\bibinfo {author} {\bibfnamefont {E.}~\bibnamefont
  {Liebhaber}}, \bibinfo {author} {\bibfnamefont {S.}~\bibnamefont
  {Acero~Gonzalez}}, \bibinfo {author} {\bibfnamefont {R.}~\bibnamefont
  {Baba}}, \bibinfo {author} {\bibfnamefont {G.}~\bibnamefont {Reecht}},
  \bibinfo {author} {\bibfnamefont {B.~W.}\ \bibnamefont {Heinrich}}, \bibinfo
  {author} {\bibfnamefont {S.}~\bibnamefont {Rohlf}}, \bibinfo {author}
  {\bibfnamefont {K.}~\bibnamefont {Rossnagel}}, \bibinfo {author}
  {\bibfnamefont {F.}~\bibnamefont {von Oppen}},\ and\ \bibinfo {author}
  {\bibfnamefont {K.~J.}\ \bibnamefont {Franke}},\ }\bibfield  {title}
  {\bibinfo {title} {{Y}u-{S}hiba-{R}usinov states in the charge-density
  modulated superconductor {N}b{S}e$_2$},\ }\href
  {https://doi.org/10.1021/acs.nanolett.9b03988} {\bibfield  {journal}
  {\bibinfo  {journal} {Nano Lett.}\ }\textbf {\bibinfo {volume} {20}},\
  \bibinfo {pages} {339} (\bibinfo {year} {2019})}\BibitemShut {NoStop}%
\bibitem [{\citenamefont {Flatt{\'e}}\ and\ \citenamefont
  {Reynolds}(2000)}]{Flatte2000}%
  \BibitemOpen
  \bibfield  {author} {\bibinfo {author} {\bibfnamefont {M.~E.}\ \bibnamefont
  {Flatt{\'e}}}\ and\ \bibinfo {author} {\bibfnamefont {D.~E.}\ \bibnamefont
  {Reynolds}},\ }\bibfield  {title} {\bibinfo {title} {Local spectrum of a
  superconductor as a probe of interactions between magnetic impurities},\
  }\href {https://doi.org/10.1103/PhysRevB.61.14810} {\bibfield  {journal}
  {\bibinfo  {journal} {Phys. Rev. B}\ }\textbf {\bibinfo {volume} {61}},\
  \bibinfo {pages} {14810} (\bibinfo {year} {2000})}\BibitemShut {NoStop}%
\bibitem [{\citenamefont {Schneider}\ \emph {et~al.}(2021)\citenamefont
  {Schneider}, \citenamefont {Beck}, \citenamefont {Posske}, \citenamefont
  {Crawford}, \citenamefont {Mascot}, \citenamefont {Rachel}, \citenamefont
  {Wiesendanger},\ and\ \citenamefont {Wiebe}}]{Schneider2021}%
  \BibitemOpen
  \bibfield  {author} {\bibinfo {author} {\bibfnamefont {L.}~\bibnamefont
  {Schneider}}, \bibinfo {author} {\bibfnamefont {P.}~\bibnamefont {Beck}},
  \bibinfo {author} {\bibfnamefont {T.}~\bibnamefont {Posske}}, \bibinfo
  {author} {\bibfnamefont {D.}~\bibnamefont {Crawford}}, \bibinfo {author}
  {\bibfnamefont {E.}~\bibnamefont {Mascot}}, \bibinfo {author} {\bibfnamefont
  {S.}~\bibnamefont {Rachel}}, \bibinfo {author} {\bibfnamefont
  {R.}~\bibnamefont {Wiesendanger}},\ and\ \bibinfo {author} {\bibfnamefont
  {J.}~\bibnamefont {Wiebe}},\ }\bibfield  {title} {\bibinfo {title}
  {Topological {S}hiba bands in artificial spin chains on superconductors},\
  }\href {https://doi.org/10.1038/s41567-021-01234-y} {\bibfield  {journal}
  {\bibinfo  {journal} {Nat. Phys.}\ }\textbf {\bibinfo {volume} {17}},\
  \bibinfo {pages} {943} (\bibinfo {year} {2021})}\BibitemShut {NoStop}%
\bibitem [{\citenamefont {Zhang}\ \emph {et~al.}(2013)\citenamefont {Zhang},
  \citenamefont {Kahle}, \citenamefont {Herden}, \citenamefont {Stroh},
  \citenamefont {Mayor}, \citenamefont {Schlickum}, \citenamefont {Ternes},
  \citenamefont {Wahl},\ and\ \citenamefont {Kern}}]{Zhang2013}%
  \BibitemOpen
  \bibfield  {author} {\bibinfo {author} {\bibfnamefont {Y.-h.}\ \bibnamefont
  {Zhang}}, \bibinfo {author} {\bibfnamefont {S.}~\bibnamefont {Kahle}},
  \bibinfo {author} {\bibfnamefont {T.}~\bibnamefont {Herden}}, \bibinfo
  {author} {\bibfnamefont {C.}~\bibnamefont {Stroh}}, \bibinfo {author}
  {\bibfnamefont {M.}~\bibnamefont {Mayor}}, \bibinfo {author} {\bibfnamefont
  {U.}~\bibnamefont {Schlickum}}, \bibinfo {author} {\bibfnamefont
  {M.}~\bibnamefont {Ternes}}, \bibinfo {author} {\bibfnamefont
  {P.}~\bibnamefont {Wahl}},\ and\ \bibinfo {author} {\bibfnamefont
  {K.}~\bibnamefont {Kern}},\ }\bibfield  {title} {\bibinfo {title}
  {Temperature and magnetic field dependence of a {K}ondo system in the weak
  coupling regime},\ }\href
  {https://doi.org/https://doi.org/10.1038/ncomms3110} {\bibfield  {journal}
  {\bibinfo  {journal} {Nat. Commun.}\ }\textbf {\bibinfo {volume} {4}},\
  \bibinfo {pages} {2110} (\bibinfo {year} {2013})}\BibitemShut {NoStop}%
\bibitem [{\citenamefont {Nagaoka}\ \emph {et~al.}(2002)\citenamefont
  {Nagaoka}, \citenamefont {Jamneala}, \citenamefont {Grobis},\ and\
  \citenamefont {Crommie}}]{Nagaoka2002}%
  \BibitemOpen
  \bibfield  {author} {\bibinfo {author} {\bibfnamefont {K.}~\bibnamefont
  {Nagaoka}}, \bibinfo {author} {\bibfnamefont {T.}~\bibnamefont {Jamneala}},
  \bibinfo {author} {\bibfnamefont {M.}~\bibnamefont {Grobis}},\ and\ \bibinfo
  {author} {\bibfnamefont {M.}~\bibnamefont {Crommie}},\ }\bibfield  {title}
  {\bibinfo {title} {Temperature dependence of a single {K}ondo impurity},\
  }\href {https://doi.org/10.1103/PhysRevLett.88.077205} {\bibfield  {journal}
  {\bibinfo  {journal} {Phys. Rev. Lett.}\ }\textbf {\bibinfo {volume} {88}},\
  \bibinfo {pages} {077205} (\bibinfo {year} {2002})}\BibitemShut {NoStop}%
\end{thebibliography}

\begin{thebibliography}{10}%
\makeatletter
\providecommand \@ifxundefined [1]{%
 \@ifx{#1\undefined}
}%
\providecommand \@ifnum [1]{%
 \ifnum #1\expandafter \@firstoftwo
 \else \expandafter \@secondoftwo
 \fi
}%
\providecommand \@ifx [1]{%
 \ifx #1\expandafter \@firstoftwo
 \else \expandafter \@secondoftwo
 \fi
}%
\providecommand \natexlab [1]{#1}%
\providecommand \enquote  [1]{``#1''}%
\providecommand \bibnamefont  [1]{#1}%
\providecommand \bibfnamefont [1]{#1}%
\providecommand \citenamefont [1]{#1}%
\providecommand \href@noop [0]{\@secondoftwo}%
\providecommand \href [0]{\begingroup \@sanitize@url \@href}%
\providecommand \@href[1]{\@@startlink{#1}\@@href}%
\providecommand \@@href[1]{\endgroup#1\@@endlink}%
\providecommand \@sanitize@url [0]{\catcode `\\12\catcode `\$12\catcode
  `\&12\catcode `\#12\catcode `\^12\catcode `\_12\catcode `\%12\relax}%
\providecommand \@@startlink[1]{}%
\providecommand \@@endlink[0]{}%
\providecommand \url  [0]{\begingroup\@sanitize@url \@url }%
\providecommand \@url [1]{\endgroup\@href {#1}{\urlprefix }}%
\providecommand \urlprefix  [0]{URL }%
\providecommand \Eprint [0]{\href }%
\providecommand \doibase [0]{https://doi.org/}%
\providecommand \selectlanguage [0]{\@gobble}%
\providecommand \bibinfo  [0]{\@secondoftwo}%
\providecommand \bibfield  [0]{\@secondoftwo}%
\providecommand \translation [1]{[#1]}%
\providecommand \BibitemOpen [0]{}%
\providecommand \bibitemStop [0]{}%
\providecommand \bibitemNoStop [0]{.\EOS\space}%
\providecommand \EOS [0]{\spacefactor3000\relax}%
\providecommand \BibitemShut  [1]{\csname bibitem#1\endcsname}%
\let\auto@bib@innerbib\@empty
%</preamble>
\bibitem [{\citenamefont {Farinacci}\ \emph {et~al.}(2018)\citenamefont
  {Farinacci}, \citenamefont {Ahmadi}, \citenamefont {Reecht}, \citenamefont
  {Ruby}, \citenamefont {Bogdanoff}, \citenamefont {Peters}, \citenamefont
  {Heinrich}, \citenamefont {von Oppen},\ and\ \citenamefont
  {Franke}}]{SFarinacci2018}%
  \BibitemOpen
  \bibfield  {author} {\bibinfo {author} {\bibfnamefont {L.}~\bibnamefont
  {Farinacci}}, \bibinfo {author} {\bibfnamefont {G.}~\bibnamefont {Ahmadi}},
  \bibinfo {author} {\bibfnamefont {G.}~\bibnamefont {Reecht}}, \bibinfo
  {author} {\bibfnamefont {M.}~\bibnamefont {Ruby}}, \bibinfo {author}
  {\bibfnamefont {N.}~\bibnamefont {Bogdanoff}}, \bibinfo {author}
  {\bibfnamefont {O.}~\bibnamefont {Peters}}, \bibinfo {author} {\bibfnamefont
  {B.~W.}\ \bibnamefont {Heinrich}}, \bibinfo {author} {\bibfnamefont
  {F.}~\bibnamefont {von Oppen}},\ and\ \bibinfo {author} {\bibfnamefont
  {K.~J.}\ \bibnamefont {Franke}},\ }\bibfield  {title} {\bibinfo {title}
  {{Tuning the coupling of an individual magnetic impurity to a superconductor:
  Quantum phase transition and transport}},\ }\href
  {https://doi.org/10.1103/PhysRevLett.121.196803} {\bibfield  {journal}
  {\bibinfo  {journal} {Phys. Rev. Lett.}\ }\textbf {\bibinfo {volume} {121}},\
  \bibinfo {pages} {196803} (\bibinfo {year} {2018})}\BibitemShut {NoStop}%
\bibitem [{\citenamefont {Heinrich}\ \emph {et~al.}(2013)\citenamefont
  {Heinrich}, \citenamefont {Braun}, \citenamefont {Pascual},\ and\
  \citenamefont {Franke}}]{SHeinrich2013}%
  \BibitemOpen
  \bibfield  {author} {\bibinfo {author} {\bibfnamefont {B.~W.}\ \bibnamefont
  {Heinrich}}, \bibinfo {author} {\bibfnamefont {L.}~\bibnamefont {Braun}},
  \bibinfo {author} {\bibfnamefont {J.~I.}\ \bibnamefont {Pascual}},\ and\
  \bibinfo {author} {\bibfnamefont {K.~J.}\ \bibnamefont {Franke}},\ }\bibfield
   {title} {\bibinfo {title} {{Protection of excited spin states by a
  superconducting energy gap}},\ }\href {https://doi.org/10.1038/nphys2794}
  {\bibfield  {journal} {\bibinfo  {journal} {Nat. Phys.}\ }\textbf {\bibinfo
  {volume} {9}},\ \bibinfo {pages} {765} (\bibinfo {year} {2013})}\BibitemShut
  {NoStop}%
\bibitem [{\citenamefont {Farinacci}\ \emph {et~al.}(2020)\citenamefont
  {Farinacci}, \citenamefont {Ahmadi}, \citenamefont {Ruby}, \citenamefont
  {Reecht}, \citenamefont {Heinrich}, \citenamefont {Czekelius}, \citenamefont
  {von Oppen},\ and\ \citenamefont {Franke}}]{SFarinacci2020}%
  \BibitemOpen
  \bibfield  {author} {\bibinfo {author} {\bibfnamefont {L.}~\bibnamefont
  {Farinacci}}, \bibinfo {author} {\bibfnamefont {G.}~\bibnamefont {Ahmadi}},
  \bibinfo {author} {\bibfnamefont {M.}~\bibnamefont {Ruby}}, \bibinfo {author}
  {\bibfnamefont {G.}~\bibnamefont {Reecht}}, \bibinfo {author} {\bibfnamefont
  {B.~W.}\ \bibnamefont {Heinrich}}, \bibinfo {author} {\bibfnamefont
  {C.}~\bibnamefont {Czekelius}}, \bibinfo {author} {\bibfnamefont
  {F.}~\bibnamefont {von Oppen}},\ and\ \bibinfo {author} {\bibfnamefont
  {K.~J.}\ \bibnamefont {Franke}},\ }\bibfield  {title} {\bibinfo {title}
  {Interfering tunneling paths through magnetic molecules on superconductors:
  Asymmetries of {K}ondo and {Y}u-{S}hiba-{R}usinov resonances},\ }\href
  {https://doi.org/10.1103/PhysRevLett.121.196803} {\bibfield  {journal}
  {\bibinfo  {journal} {Phys. Rev. Lett.}\ }\textbf {\bibinfo {volume} {125}},\
  \bibinfo {pages} {256805} (\bibinfo {year} {2020})}\BibitemShut {NoStop}%
\bibitem [{\citenamefont {Pillet}\ \emph {et~al.}(2010)\citenamefont {Pillet},
  \citenamefont {Quay}, \citenamefont {Morfin}, \citenamefont {Bena},
  \citenamefont {Yeyati},\ and\ \citenamefont {Joyez}}]{SPillet2010}%
  \BibitemOpen
  \bibfield  {author} {\bibinfo {author} {\bibfnamefont {J.}~\bibnamefont
  {Pillet}}, \bibinfo {author} {\bibfnamefont {C.}~\bibnamefont {Quay}},
  \bibinfo {author} {\bibfnamefont {P.}~\bibnamefont {Morfin}}, \bibinfo
  {author} {\bibfnamefont {C.}~\bibnamefont {Bena}}, \bibinfo {author}
  {\bibfnamefont {A.~L.}\ \bibnamefont {Yeyati}},\ and\ \bibinfo {author}
  {\bibfnamefont {P.}~\bibnamefont {Joyez}},\ }\bibfield  {title} {\bibinfo
  {title} {Andreev bound states in supercurrent-carrying carbon nanotubes
  revealed},\ }\href {https://doi.org/10.1038/nphys1811} {\bibfield  {journal}
  {\bibinfo  {journal} {Nat. Phys.}\ }\textbf {\bibinfo {volume} {6}},\
  \bibinfo {pages} {965} (\bibinfo {year} {2010})}\BibitemShut {NoStop}%
\bibitem [{\citenamefont {Malavolti}\ \emph {et~al.}(2018)\citenamefont
  {Malavolti}, \citenamefont {Briganti}, \citenamefont {H{\"{a}}nze},
  \citenamefont {Serrano}, \citenamefont {Cimatti}, \citenamefont {McMurtrie},
  \citenamefont {Otero}, \citenamefont {Ohresser}, \citenamefont {Totti},
  \citenamefont {Mannini}, \citenamefont {Sessoli},\ and\ \citenamefont
  {Loth}}]{SMalavolti2018}%
  \BibitemOpen
  \bibfield  {author} {\bibinfo {author} {\bibfnamefont {L.}~\bibnamefont
  {Malavolti}}, \bibinfo {author} {\bibfnamefont {M.}~\bibnamefont {Briganti}},
  \bibinfo {author} {\bibfnamefont {M.}~\bibnamefont {H{\"{a}}nze}}, \bibinfo
  {author} {\bibfnamefont {G.}~\bibnamefont {Serrano}}, \bibinfo {author}
  {\bibfnamefont {I.}~\bibnamefont {Cimatti}}, \bibinfo {author} {\bibfnamefont
  {G.}~\bibnamefont {McMurtrie}}, \bibinfo {author} {\bibfnamefont
  {E.}~\bibnamefont {Otero}}, \bibinfo {author} {\bibfnamefont
  {P.}~\bibnamefont {Ohresser}}, \bibinfo {author} {\bibfnamefont
  {F.}~\bibnamefont {Totti}}, \bibinfo {author} {\bibfnamefont
  {M.}~\bibnamefont {Mannini}}, \bibinfo {author} {\bibfnamefont
  {R.}~\bibnamefont {Sessoli}},\ and\ \bibinfo {author} {\bibfnamefont
  {S.}~\bibnamefont {Loth}},\ }\bibfield  {title} {\bibinfo {title} {Tunable
  spin-superconductor coupling of spin $1/2$ vanadyl phthalocyanine
  molecules},\ }\href {https://doi.org/10.1021/acs.nanolett.8b03921} {\bibfield
   {journal} {\bibinfo  {journal} {Nano Lett.}\ }\textbf {\bibinfo {volume}
  {18}},\ \bibinfo {pages} {7955} (\bibinfo {year} {2018})}\BibitemShut
  {NoStop}%
\bibitem [{\citenamefont {Frank}\ and\ \citenamefont
  {Jacob}(2015)}]{SFrank2015}%
  \BibitemOpen
  \bibfield  {author} {\bibinfo {author} {\bibfnamefont {S.}~\bibnamefont
  {Frank}}\ and\ \bibinfo {author} {\bibfnamefont {D.}~\bibnamefont {Jacob}},\
  }\bibfield  {title} {\bibinfo {title} {Orbital signatures of {Fano-Kondo}
  line shapes in {STM} adatom spectroscopy},\ }\href
  {https://doi.org/10.1103/PhysRevB.92.235127} {\bibfield  {journal} {\bibinfo
  {journal} {Phys. Rev. B}\ }\textbf {\bibinfo {volume} {92}},\ \bibinfo
  {pages} {235127} (\bibinfo {year} {2015})}\BibitemShut {NoStop}%
\bibitem [{Note1()}]{Note1}%
  \BibitemOpen
  \bibinfo {note} {The symmetry of the hybridization will in general depend on
  the YSR wave function of the monomer \cite {SRuby2016}.}\BibitemShut {Stop}%
\bibitem [{\citenamefont {Pientka}\ \emph {et~al.}(2013)\citenamefont
  {Pientka}, \citenamefont {Glazman},\ and\ \citenamefont {von
  Oppen}}]{SPientka2013}%
  \BibitemOpen
  \bibfield  {author} {\bibinfo {author} {\bibfnamefont {F.}~\bibnamefont
  {Pientka}}, \bibinfo {author} {\bibfnamefont {L.~I.}\ \bibnamefont
  {Glazman}},\ and\ \bibinfo {author} {\bibfnamefont {F.}~\bibnamefont {von
  Oppen}},\ }\bibfield  {title} {\bibinfo {title} {{Topological superconducting
  phase in helical Shiba chains}},\ }\href
  {https://doi.org/10.1103/PhysRevB.88.155420} {\bibfield  {journal} {\bibinfo
  {journal} {Phys. Rev. B}\ }\textbf {\bibinfo {volume} {88}},\ \bibinfo
  {pages} {155420} (\bibinfo {year} {2013})}\BibitemShut {NoStop}%
\bibitem [{\citenamefont {R\"ontynen}\ and\ \citenamefont
  {Ojanen}(2015)}]{SOjanen2015}%
  \BibitemOpen
  \bibfield  {author} {\bibinfo {author} {\bibfnamefont {J.}~\bibnamefont
  {R\"ontynen}}\ and\ \bibinfo {author} {\bibfnamefont {T.}~\bibnamefont
  {Ojanen}},\ }\bibfield  {title} {\bibinfo {title} {{Topological
  Superconductivity and High Chern Numbers in 2D Ferromagnetic Shiba
  Lattices}},\ }\href {https://doi.org/10.1103/PhysRevLett.114.236803}
  {\bibfield  {journal} {\bibinfo  {journal} {Phys. Rev. Lett.}\ }\textbf
  {\bibinfo {volume} {114}},\ \bibinfo {pages} {236803} (\bibinfo {year}
  {2015})}\BibitemShut {NoStop}%
\bibitem [{\citenamefont {Ruby}\ \emph {et~al.}(2016)\citenamefont {Ruby},
  \citenamefont {Peng}, \citenamefont {von Oppen}, \citenamefont {Heinrich},\
  and\ \citenamefont {Franke}}]{SRuby2016}%
  \BibitemOpen
  \bibfield  {author} {\bibinfo {author} {\bibfnamefont {M.}~\bibnamefont
  {Ruby}}, \bibinfo {author} {\bibfnamefont {Y.}~\bibnamefont {Peng}}, \bibinfo
  {author} {\bibfnamefont {F.}~\bibnamefont {von Oppen}}, \bibinfo {author}
  {\bibfnamefont {B.~W.}\ \bibnamefont {Heinrich}},\ and\ \bibinfo {author}
  {\bibfnamefont {K.~J.}\ \bibnamefont {Franke}},\ }\bibfield  {title}
  {\bibinfo {title} {{Orbital Picture of Yu-Shiba-Rusinov Multiplets}},\ }\href
  {https://doi.org/10.1103/PhysRevLett.117.186801} {\bibfield  {journal}
  {\bibinfo  {journal} {Phys. Rev. Lett.}\ }\textbf {\bibinfo {volume} {117}},\
  \bibinfo {pages} {186801} (\bibinfo {year} {2016})}\BibitemShut {NoStop}%
\end{thebibliography}
%

\clearpage

\setcounter{figure}{0}
\setcounter{section}{0}
\setcounter{equation}{0}
\setcounter{table}{0}
\renewcommand{\theequation}{S\arabic{equation}}
\renewcommand{\thefigure}{S\arabic{figure}}
	\renewcommand{\thetable}{S\arabic{table}}%
	\setcounter{section}{0}
	\renewcommand{\thesection}{S\arabic{section}}%

\onecolumngrid

\maketitle 
\section*{Supplementary Material}
\section{Identification of de-chlorinated FeP molecules}

In the main text, we described the self-assembly of de-chlorinated FeP molecules and Cl atoms. We argued that the Cl atoms detached from the FeP-Cl molecules upon adsorption, and that the amount of Cl atoms on the surface, and thus the amount, which is available to participating in the self-assembly, can be tuned by the annealing temperature. Yet, we found that the dechlorinated molecules can appear with two slightly different apparent heights in the STM images. The brighter molecules (clover shape) may - at first sight - be interpreted as the chlorinated species. While our systematic analysis with increasing temperature as described in the main text already suggests that both types of molecules are de-chlorinated, we provide another point of evidence that the types "only" exhibit different electronic properties here.

\begin{figure*}[b]
    \centering
    \includegraphics[width=0.9\linewidth]{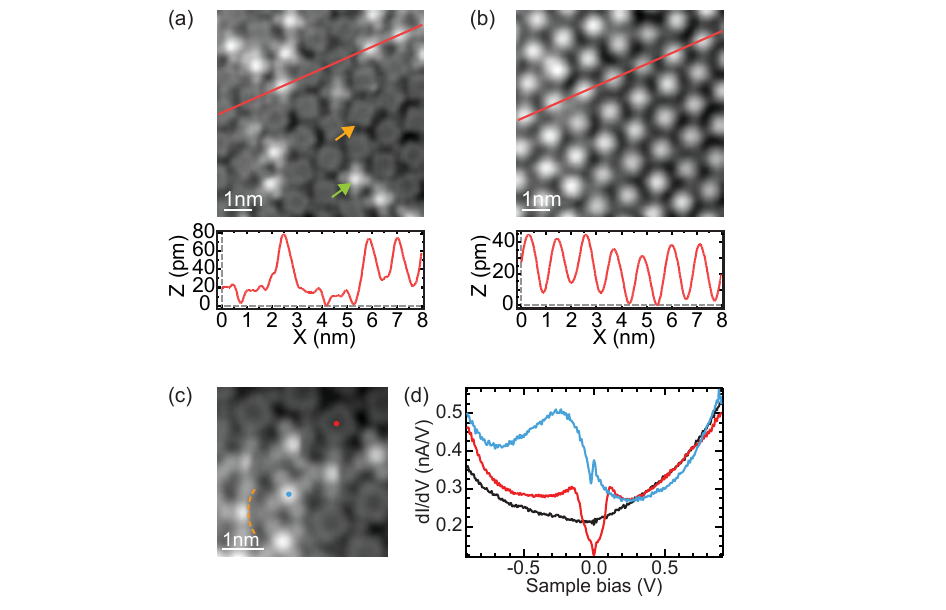}
    \caption{ (a) Topography image recorded with $V_\mathrm{bias}=5$~mV, $I=200$~pA in which two types of molecules can be identified (see green and orange arrows), as exemplified by the height profile taken along the red line. (b) Topography image of the same area recorded with $V_\mathrm{bias}=700$~mV, $I=200$~pA: the molecules now have the same appearance as illustrated by the corresponding line profile. (c) Topography images ($V_\mathrm{bias}=5$~mV, $I=200$~pA) where the two types of molecules can be identified -- a few molecules display both appearances (see orange dashed line). The dots indicate the locations of the \didv\ spectra in (c). (c) \didv\ spectra recorded above the centers of a clover-shaped molecule (blue) and a square-shaped molecule (red) along with a reference on bare Pb (black). Feedback opened at $V_\mathrm{bias}=900$~mV, $I=300$~pA and signal modulated with $V_\mathrm{rms}=5$~mV.
    }
    \label{fig:DarkBright}
\end{figure*}

In Fig.\ \ref{fig:DarkBright}a we show an STM image recorded at low bias voltage, where two types of molecules appear with different height (clover shape with a central protrusion and square-like). An image taken at larger bias voltage of the same area reveals all molecules with identical appearance (Fig.\ \ref{fig:DarkBright}b), suggesting that none of the molecules carries a Cl ligand attached to the Fe center. The different electronic structure leading to different appearance in the STM image is also reflected in their different \didv\ spectra. Molecules, which appear clover-shape and bright at low bias, exhibit a broad resonance around -300~mV, whereas the dark square-like molecules exhibit a broad pair of symmetric steps around the Fermi energy. Note also that in Fig.\ \ref{fig:DarkBright}c some molecules have a hybrid appearance: clover-shaped on the left and central part, square-shaped on the right (the orange dashed line serves as a guide to the eye). We speculate that the presence of Cl adatoms effectively gates the electronic properties of the FeP molecules in their vicinity, which leads to these two different molecular types.

In very rare cases, we also observe molecules with a very bright protrusion above their center ($<1$\%), as shown in the supplementary information of \citep{SFarinacci2018}. They exhibit two pairs of resonances outside the superconducting gap in \didv\ spectra, which are a fingerprint of spin excitations for magnetic adsorbates on superconductors. As a matter of fact, their \didv\ spectra is almost identical to that of  iron-octa-ethyl-porphyrin-chloride (FeOEP-Cl) molecules on Pb(111) \cite{SHeinrich2013}. FeOEP and FeP molecules only differ by the presence of additional ligands for FeOEP, away from the Fe center. Since the oxidation state, spin state and magnetic anisotropy of Fe is mostly dictated by its direct surroundings, we conclude that the molecules with a very high protrusion have kept their Cl ligand upon adsorption. The two types of molecules presented in the main text are thus assigned to FeP molecules, and the small protrusions in between them to Cl adatoms.

\section{YSR states at the edge of kagome precursors}

\begin{figure*}[ht!]
    \centering
    \includegraphics[width=0.9\linewidth]{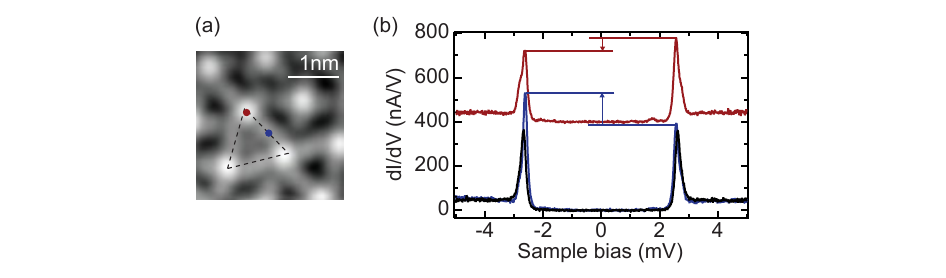}
    \caption{(a) Topography image of a single tile made of three FeP molecules and one Cl adatom ($V_\mathrm{bias}=5$~mV, $I=200$~pA). (b) \didv\ spectra taken above the center (red -- offset for clarity) and ligand (blue) of a FeP molecule as indicated in (a), along with a reference spectrum on bare Pb (black). Feedback opened at $V_\mathrm{bias}=5$~mV, $I=200$~pA and signal modulated with $V_\mathrm{rms}=15~\mu$eV.
    }
    \label{fig:YSREdge}
\end{figure*}

In the main paper we show evidence of hybridization between YSR states induced by molecules surrounded by two Cl adatoms. The presence of these two adatoms is crucial for the YSR hybridization. Indeed, when there is only one Cl adatom in the vicinity, the FeP molecules do not display any YSR state well inside the gap but rather an asymmetry in the intensity of the coherence peaks (Fig. 3b of the main text and Fig.\ \ref{fig:YSREdge}). The opposite intensity asymmetry above the Fe center and molecular ligand is consistent with the presence of a YSR state close to the gap edge with the asymmetry being a result of different interfering tunneling paths \cite{SFarinacci2020}. The absence of splitting in the \didv\ spectra indicates that, if any hybridzation takes place between molecules surrounded by only one Cl adatom, it remains below our energy resolution and at a different energy that those that pertain to the YSR hybridization described in the main paper.

\section{Deconvolution of \didv\ spectra and fitting of the sample LDOS}

\begin{figure*}[ht!]
    \centering
    \includegraphics[width=0.9\linewidth]{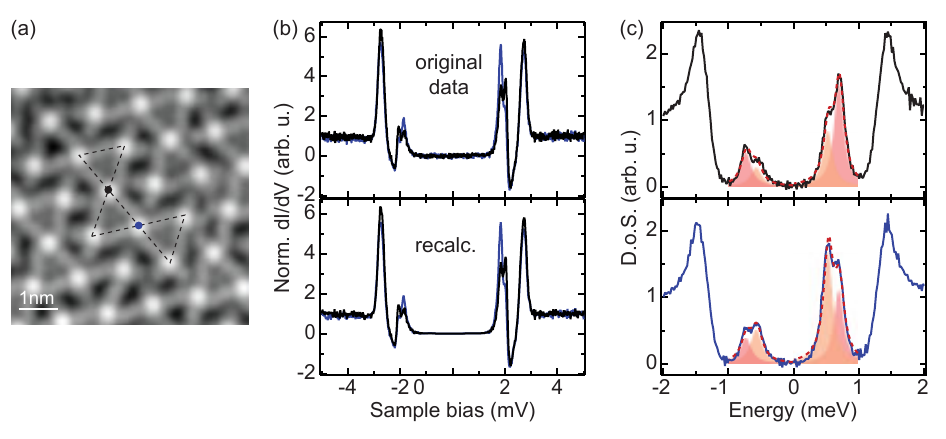}
    \caption{(a) Topography image of a kagome precursor ($V_\mathrm{bias}=5$~mV, $I=200$~pA). (b) Upper graph: \didv\ spectra taken at the locations indicated in (a) (feedback opened at $V_\mathrm{bias}=5$~mV, $I=200$~pA and signal modulated with $V_\mathrm{rms}=15~\mu$eV). Lower graph: reconvolved spectra. The match between the two graphs ensures the consistency of the deconvolution procedure. (c) Sample DOS obtained by deconvolving the \didv\ spectra in (b). Red dashed lines show fits with a sum of Lorentzian peaks for both bias polarities, the contribution of each Lorentzian being indicated by filled curves. 
    }
    \label{fig:Deconvolution}
\end{figure*}

In order to investigate the distribution of the hybridized YSR states it is important to remove the influnce of the density of states (DOS) of the tip. Indeed, the non-flat DOS of the superconducting tip can lead to negative differential resistance effects that would artificially lower the intensity of a YSR state in close vicinity of another one.

We perform a numerical deconvolution of the \didv\ spectra following the approach of Pillet et al.\cite{SPillet2010}. Neglecting proportionality constants, the tunneling current is given by:

\begin{align}
I = \int [f(\omega - eV) - f(\omega)]\rho_t(\omega-eV)\rho_s(\omega) d\omega,
\end{align}
where $f$ is the Fermi function, $\rho_s$ ($\rho_t$) the DOS of the sample (tip) and $V$ the bias voltage. As a result, the differential conductance is the sum of two integrals.

\begin{equation}
\frac{dI}{dV} = \int - e[f(\omega - eV)-f(\omega)] \frac{\partial \rho_t}{\partial E}\bigg|_{E=\omega - eV}\rho_s(\omega) d\omega +
\int -e \frac{\partial f}{\partial E} \bigg|_{E=\omega - eV} \rho_t(\omega - eV) \rho_s(\omega)d\omega.
\end{equation}

These two integrals are of the form $\int K(\omega)\rho_s(\omega)d\omega$ where

\begin{align}
K_1(\omega) &= -e[f(\omega-eV)-f(\omega)]\frac{\partial \rho_t}{\partial E}\bigg|_{E=\omega - eV} \\
K_2(\omega) &= -e\frac{\partial f}{\partial E} \bigg|_{E=\omega - eV} \rho_t(\omega - eV).
\end{align}

They can be numerically approximated via a discretization of the energy and bias voltage so that the differential conductance is simply given by a matrix product:

\begin{equation}
\frac{dI}{dV	} = K \cdot \rho_s,
\end{equation}

where $K=K_1+K_2$ depends only on the tip DOS that we assume to be Bardeen-Cooper-Schrieffer (BCS) like.

The sample DOS is then obtained by pseudo-inverting the matrix $K$:

\begin{equation}
\rho_s = K^{-1}\cdot \frac{dI}{dV}.
\end{equation}

To ensure the consistency of the deconvolution procedure, we reconvolve the resulting DOS with the $K$ matrix and verify that the obtained spectrum corresponds to the original \didv\ spectrum.
Two examples of this are shown in Fig.\ \ref{fig:Deconvolution}. Subsequently, we fit the sample DOS with a sum of Lorentzians. The amplitude of each Lorentzian indicates the amplitude of the corresponding YSR state at the position where the spectrum was taken. In the main text, we only display the amplitude of the fits at positive energies, but we ensured that the negative-energy results show the same patterns.

\section{Identification of the YSR ground state}

\begin{figure*}[ht!]
    \centering
    \includegraphics[width=0.9\linewidth]{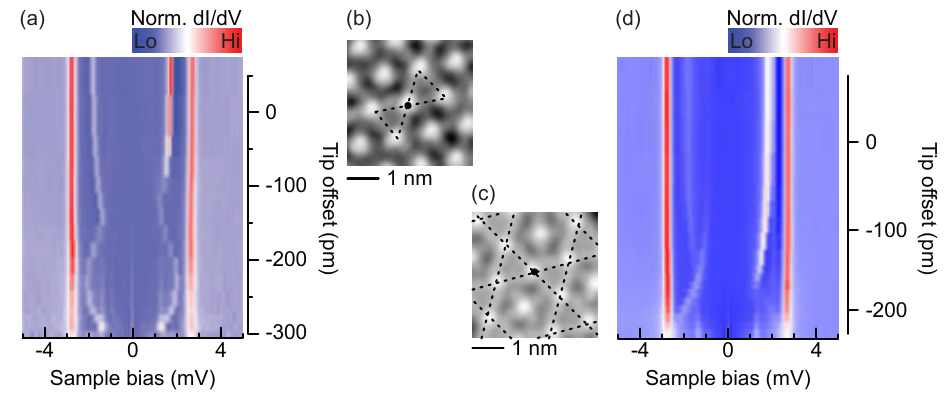}
    \caption{(a) \didv\ spectra measured above the center of a Kagome precursor as shown in (b) ($V_\textrm{bias} = 5$~mV, $I=200$~pA) as a function of tip offset after opening the feedback with $V_\textrm{bias} = 5$~mV, $I = 200$~pA (negative offsets defined as bringing the tip closer to the surface, signal modulated with $V_\textrm{rms} = 15\mu $~eV). (c) Topography image ($V_\textrm{bias} = 5$~mV, $I= 200$~pA) in which a kagome lattice is formed, as indicated by black dashed lines. A black circle indicates the location where the data set of (d) has been recorded. (d) \didv\ spectra as a function of tip offset after opening the feedback at $V_\textrm{bias}=5$~mV, $I=200$~pA (signal modulated with $V_\textrm{rms}= 20 \mu $eV).
    }
    \label{fig:Approach}
\end{figure*}

In order to identify the ground state of the system we perform a tip approach over the Fe center of a FeP molecule \cite{SFarinacci2018, SMalavolti2018}. We perform such tip approach above a molecule that is surrounded by two Cl adatoms but without YSR hybridization (Fig.\ \ref{fig:Approach}a-b) as well as above a molecule that is inside a Kagome domain (Fig.\ \ref{fig:Approach}c-d). In both cases, the results are in line with those previously obtained above a molecule at the edge of a kagome lattice \cite{SFarinacci2018}.
As the tip is brought closer to the Fe center, the YSR states shifts towards Fermi energy and we observe a crossing of the quantum phase transition for tip offsets between $-100$~pm and $-200$~pm, from there on the YSR state shifts away from the Fermi energy. This shift of the YSR state is attributed to a weakening of the exchange coupling between the Fe center and substrate as the molecule is pulled toward the tip due to attractive van-der-Waals interactions. Going even closer in Fig.\ \ref{fig:Approach}a, we observe a reversed trend starting from $\Delta z \sim 240$~pm, with the YSR shifting toward the Fermi energy. This corresponds to the onset of the repulsive regime where the molecule is pushed back toward the Pb substrate.

All in all, these results, with in particular the shift toward the Fermi energy at large tip-sample distance, indicate that the YSR state of the molecules is in the screened regime. Note that for the experiment shown in Fig.\ \ref{fig:Approach}c-d, the YSR state is gradually detuned from its neighbours with diminishing hybridization upon tip approach.

\section{Kondo effect in the normal state of Pb}

\begin{figure*}[ht!]
    \centering
    \includegraphics[width=0.9\linewidth]{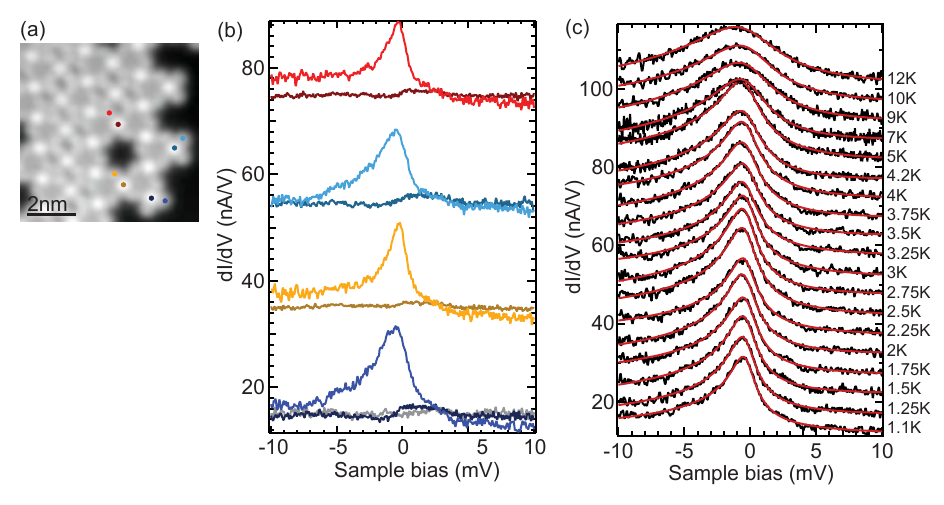}
    \caption{(a) Topography image of the molecules investigated in Fig.6 of the main text ($V_\textrm{bias} = -45$~mV, $I=30$~pA). (b) \didv\ spectra taken at the location indicated in (a) (feedback opened at $V_\textrm{bias}=10$~mV, $I=150$~pA and signal modulated with $V_\textrm{rms}=50~\mu $eV). A reference spectrum taken above bare Pb is shown in grey, spectra are offset for clarity. (c) Evolution with temperature of the Kondo resonance above the ligand of teh lowest molecule in a (dark blue) with Fano-Frota fits in red.
    }
    \label{fig:Tdep}
\end{figure*}

\begin{figure*}[ht!]
    \centering
    \includegraphics[width=0.9\linewidth]{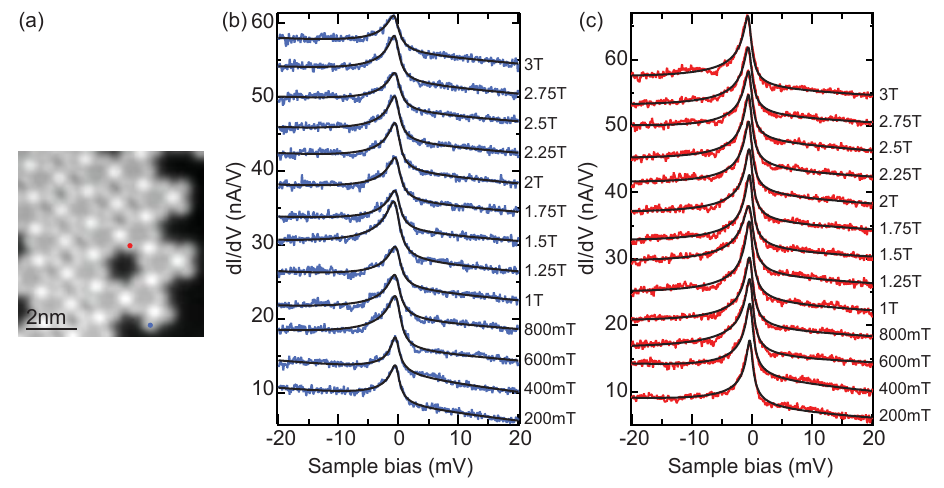}
    \caption{(a) Topography image ($V_\textrm{bias} = -45$~mV, $I=30$~pA) of the island shown in Fig.6 of the main text with the two locations where the magnetic field dependent measurements where taken. The two data sets along with the Fano-Frota fits (black curves) are displayed in (b) and (c) (feedback opened at $V_\textrm{bias}=10$~mV, $I=150$~pA and signal modulated with $V_\textrm{rms}50~\mu $eV).
    }
    \label{fig:Bfield}
\end{figure*}

In the normal state of Pb, the FeP molecules display a Kondo resonance. In Fig.\ \ref{fig:Tdep}b, we show the \didv\ spectra taken above the center and molecular ligand of the four molecules of Fig.6 of the main text (see Fig.\ \ref{fig:Tdep}a).

Due to interferences between tunneling paths \cite{SFarinacci2020}, the lineshape of the Kondo resonance is very different between the center and ligand of the molecules. The resonances are more intense and better resolved above the molecular ligands. We therefore use these positions to study the evolution of the resonance with temperature (Fig.\ \ref{fig:Tdep}) and external magnetic field (Fig.\ \ref{fig:Bfield}). In Fig.\ \ref{fig:Tdep}c we show an exemplary set of data taken at various temperatures above the molecule at the edge of the island in Fig.\ \ref{fig:Tdep}a, indicated by a dark blue circle. Each spectrum is fitted with a Fano-Frota function (see red lines) to extract the intrinsic width of the resonance \cite{SFrank2015}. Similar data sets were taken at the ligand positions of other molecules marked in the topography image of Fig.\ \ref{fig:Tdep}a. The fit results are presented in Fig.6b of the main text, where the error bars correspond to the standard deviation of the fits.

In Fig.\ \ref{fig:Bfield} we show the evolution of the Kondo resonance with external magnetic field for two molecules, one at the edge and one inside the island, as indicated in the topography image of Fig.\ \ref{fig:Bfield}a. For both molecules, the linewidth -- extracted by Fano-Frota fits shown in black -- stays constant over the range of external magnetic fields achievable in our experiment.

Overall, these results indicate a smaller coupling to the underlying substrate of the molecules inside the island than those at the edge. This is in line with the fact that these molecules exhibit YSR states at different energies in the superconducting state of the sample. Kondo width and magnetic field dependence show that the molecules lie in the strong-coupling Kondo regime with a critical field above 3~T. 

The temperature and B-field behavior are in agreement with a single spin coupled to the substrate. This together with the absence of any deviation from a Frota-Fano lineshape suggests that any Kondo-lattice behavior is weak and beyond our resolution. 

\section{Kagome lattice of Yu-Shiba-Rusinov states}

The tight-binding band structure of a kagome lattice in the normal state contains a flat band. Here, we consider how pairing effects modify this flat band in the band structure of YSR states originating from a kagome lattice of magnetic adatoms on a superconductor. We assume that the adatom spins are polarized, choosing a ferromagnetic arrangement for simplicity. Without spin-orbit coupling and assuming $s$-wave pairing of the substrate superconductor, hybridization of the subgap YSR states yields uncoupled sets of electron and hole bands. Each of these contains a flat band provided that the hybridization of YSR states is isotropic \footnote{The symmetry of the hybridization will in general depend on the YSR wave function of the monomer \cite{SRuby2016}.}. Spin-orbit coupling introduces $p$-wave pairing \cite{SPientka2013,SOjanen2015}. Assuming Rashba coupling, this pairing is of chiral $p$-wave nature. Restricting to nearest-neighbor hybridization of the YSR states, the Bogoliubov-de Gennes Bloch Hamiltonian takes the form 
\begin{equation}
     H(\mathbf{k}) = \left(\begin{array}{cc} H_n(\mathbf{k}) & \Delta(\mathbf{k})\\ \Delta^\dagger (\mathbf{k}) & - [H_n(-\mathbf{k})]^T \end{array}
     \right).
\end{equation}
Here, the normal-state Hamiltonian is
\begin{equation}
     H_n(\mathbf{k}) = \left( \begin{array}{ccc}
      \epsilon_\mathrm{YSR} & -t(1+e^{-i\mathbf{k}\cdot\mathbf{a}_1}) & 
      -t(1+e^{i\mathbf{k}\cdot\mathbf{a}_2})  \\
      -t(1+e^{i\mathbf{k}\cdot\mathbf{a}_1}) & \epsilon_\mathrm{YSR} &
      -t(1+e^{-i\mathbf{k}\cdot\mathbf{a}_3})  \\
      -t(1+e^{-i\mathbf{k}\cdot\mathbf{a}_2})  &
      -t(1+e^{i\mathbf{k}\cdot\mathbf{a}_3})  & \epsilon_\mathrm{YSR} \end{array}
     \right)
\end{equation}
with the energy $\epsilon_\mathrm{YSR}$ of the YSR state of the monomers, (isotropic) hopping $t$,  and the vectors $\mathbf{a}_1 = a [0,1]$, $\mathbf{a}_2 = a[\cos(2\pi/3),\sin(2\pi/3)]$, and $\mathbf{a}_3 = a[\cos(2\pi/3),-\sin(2\pi/3)]$ in terms of the bond length $a$. The pairing matrix takes the form
\begin{equation}
     \Delta(\mathbf{k}) = \left( \begin{array}{ccc}
      0 & \Delta e^{-i2\pi/3} (1-e^{-i\mathbf{k}\cdot\mathbf{a}_1}) & 
      \Delta e^{i2\pi/3} (1-e^{i\mathbf{k}\cdot\mathbf{a}_2})  \\
      \Delta e^{i2\pi/3} (1-e^{i\mathbf{k}\cdot\mathbf{a}_1}) & 0 &
      \Delta (1-e^{-i\mathbf{k}\cdot\mathbf{a}_3})  \\
      \Delta e^{-i2\pi/3} (1-e^{-i\mathbf{k}\cdot\mathbf{a}_2})  &
      \Delta (1-e^{i\mathbf{k}\cdot\mathbf{a}_3})  &  0 \end{array}
     \right)
\end{equation}
with induced pairing strength $\Delta$. 

Diagonalizing $H(\mathbf{k})$, one finds that even for isotropic hybridization, pairing generally gives a finite dispersion to the flat bands. The magnitude of the dispersion depends on the ratio of $\Delta/t$. In experiment, the pairing strength $\Delta$ is controlled by the strength of spin-orbit coupling. Thus, the YSR band structure will contain essentially flat electron and hole bands in the limit of weak spin-orbit coupling, while the flat-band nature is progressively lost with increasing  spin-orbit coupling. We expect these conclusions to remain valid beyond the simplifying assumptions (ferromagnetic order, nearest-neighbor hybridization, single pair of YSR resonances) made here.

\end{document}